\documentclass[letterpaper]{article} 
\usepackage{aaai25}  
\usepackage{times}  
\usepackage{helvet}  
\usepackage{courier}  
\usepackage[hyphens]{url}  
\usepackage[T1]{fontenc}
\usepackage{graphicx} 
\usepackage{natbib}  
\usepackage{caption} 
\usepackage{listings}
\usepackage{algorithm}
\usepackage{algorithmic}
\usepackage{booktabs}
\usepackage{amsmath}
\usepackage[table]{xcolor}
\usepackage{xcolor}
\usepackage{soul}
\usepackage{etoolbox}

\usepackage{tabularx}
\usepackage{makecell}
\usepackage{threeparttable}
\usepackage{newfloat}
\usepackage{listings}
\DeclareCaptionStyle{ruled}{labelfont=normalfont,labelsep=colon,strut=off} 
\floatstyle{ruled}
\newfloat{listing}{tb}{lst}{}
\floatname{listing}{Listing}
\usepackage{xcolor}
\newcommand{\answerYes}[1]{\textcolor{blue}{#1}} 
\newcommand{\answerNo}[1]{\textcolor{teal}{#1}} 
\newcommand{\answerNA}[1]{\textcolor{gray}{#1}}

\title{Quantifying the Spread of Online Incivility in Brazilian Politics}

\author {
    Yuan Zhang\textsuperscript{\rm 1},
    Michael Amsler\textsuperscript{\rm 1},
    Laia Castro\textsuperscript{\rm 2,3},
    Frank Esser\textsuperscript{\rm 1},
    Alexandre Bovet\textsuperscript{\rm 4,5}
}

\affiliations {
    \textsuperscript{\rm 1}Department of Communication and Media Research, University of Zurich, Zurich, Switzerland\\
    \textsuperscript{\rm 2}Department of Political Science, University of Barcelona, Barcelona, Spain\\
    \textsuperscript{\rm 3} Barcelona Supercomputing Center, Barcelona, Spain\\ 
    \textsuperscript{\rm 4}Department of Mathematical Modeling and Machine Learning, University of Zurich, Zurich, Switzerland\\
    \textsuperscript{\rm 5}Digital Society Initiative, University of Zurich, Zurich, Switzerland\\
    y.zhang@ikmz.uzh.ch, m.amsler@ikmz.uzh.ch,
    lcastro@ub.edu,
    f.esser@ikmz.uzh.ch, alexandre.bovet@uzh.ch
}

\begin{document}

\maketitle

\begin{abstract}
Incivility refers to behaviors that violate collective norms and disrupt cooperation within the political process. Although large-scale online data and automated techniques have enabled the quantitative analysis of uncivil discourse, prior research has predominantly focused on impoliteness or toxicity, often overlooking other behaviors that undermine democratic values. To address this gap, we propose a multidimensional conceptual framework encompassing Impoliteness (IMP), Physical Harm and Violent Political Rhetoric (PHAVPR), Hate Speech and Stereotyping (HSST), and Threats to Democratic Institutions and Values (THREAT). Using this framework, we measure the spread of online political incivility in Brazil using approximately 5 million tweets posted by 2,307 political influencers during the 2022 Brazilian general election. Through statistical modeling and network analysis, we examine the dynamics of uncivil posts at different election stages, identify key disseminators and audiences, and explore the mechanisms driving the spread of uncivil information online. Our findings indicate that impoliteness is more likely to surge during election campaigns. In contrast, the other dimensions of incivility are often triggered by specific violent events. Moreover, we find that left-aligned individual influencers are the primary disseminators of online incivility in the Brazilian Twitter/X sphere and that they disseminate not only direct incivility but also indirect incivility when discussing or opposing incivility expressed by others. They relay those content from politicians, media agents, and individuals to reach broader audiences, revealing a diffusion pattern mixing the direct and two-step flows of communication theory. This study offers new insights into the multidimensional nature of incivility in Brazilian politics and provides a conceptual framework that can be extended to other political contexts.
\end{abstract}

\section{Introduction}
Social media platforms have enabled diverse political influencers—such as politicians, media outlets, journalists, and ordinary users—to produce political content and shape public opinion \cite{goodwin2023political}. While social media creates an environment for ordinary users to receive political information by bypassing traditional information sources, it also provides a space for uncivil expressions and hostile attacks to unfold \cite{heseltine2022online}. Previous studies have shown evidence that such uncivil behavior might reduce public political trust \cite{mutz2005new}, increase polarization among groups \cite{kim2019incivility}, and even lead to offline violence \cite{gallacher2021online}. However, incivility is not always detrimental to democracy. Studies have also shown that incivility can also draw public attention \cite{mutz2005new}, encourage political participation \cite{brooks2007beyond}, and assist marginalized groups in expressing disagreement \cite{lozano2009uncivil}. Therefore, it is neither possible nor necessarily desirable to remove all uncivil content online \cite{chen2019incivility}. This realization motivates many studies, including this one, to achieve a deeper understanding of incivility by measuring it through more nuanced categories \cite{gao2024crisis, stryker2016political, muddiman2017personal, bentivegna2022searching, papacharissi2004democracy}. 

Incivility has been previously subcategorized into behaviors such as name-calling, aspersion, vulgarity, hyperbole, and shouting/screaming notation, which involve disrespectful or rude actions towards individuals or groups \cite{coe2014online, otto2020context, mutz2007effects}. Others argue that certain discourses may remain polite yet still be uncivil if only they violate collective norms established, such as  discourse that undermines democratic values \cite{gao2024crisis, papacharissi2004democracy, rossini2022beyond}. While many discussions about the multidimensional concept of incivility exist \cite{muddiman2017personal, bentivegna2022searching}, there remains a lack of relevant measurements and deeper understanding regarding who the main disseminators and audiences are and how such content spreads among users.

In this study, we propose a multidimensional conceptual framework that synthesizes prior research, classifying incivility based on violations of collective norms such as social norms and democratic norms, along with four specific taxonomies: Impoliteness (IMP), Physical Harm and Violent Political Rhetoric (PHAVPR), Hate Speech and Stereotyping (HSST), and Threats to Democratic Institutions and Values (THREAT). We then develop algorithms to facilitate the detection of these four dimensions of incivility.

Our ultimate goal is threefold: first, to understand the dynamics of different dimensions of incivility online, focusing on trends and key events. Second, to identify the main disseminators and audiences of incivility. Finally, to investigate the mechanisms of information flow for online incivility. In summary, this study is organized around the following research questions:\\

\noindent\textbf{RQ1:} How can we construct multidimensional incivility, and how can we measure it?\\ 
\textbf{RQ2:} When do different dimensions of online incivility emerge?\\ 
\textbf{RQ3:} Who are the main disseminators and audiences of different dimensions of online incivility?\\ 
\textbf{RQ4:} In what ways do political influencers disseminate uncivil content to ordinary audiences?\\

To address these questions, we combine national survey data collected during the 2022 Brazilian Presidential Election—held in two rounds on October 2 (first round) and October 30 (runoff)—with 5 million posts from 2,307 Brazilian political influencers on Twitter (now X) followed by survey participants. First, for RQ1, we develop a codebook that distinguishes incivility dimensions based on violations of different collective norms and train human coders for manual annotation. We then develop multiple binary classifiers using pre-trained sentence transformers and apply them for classification. For RQ2, we employ non-parametric modeling to fit the trends in the frequency of uncivil content and detect key events. For RQ3, we manually annotate political influencers into three account types—Politician, Media, and Individual—and categorize their socio-political identities. We investigate their distributions across incivility dimensions. Finally, for RQ4, we construct an incivility network using following and retweeting relations. We detect three network motifs representing different information flow mechanisms among survey users and political influencers. We also use centrality measures in retweet networks of uncivil messages to identify the main original creators of incivility.\\

Our analyses show that scores for all dimensions of incivility peak during heated events and decrease subsequently. However, IMP primarily breaks out before the election days. PHAVPR, HSST, and THREAT mainly emerge during or after events of extreme-right violence. Second, a deeper dive into the disseminators reveals that left-aligned individual political influencers constitute the majority of those disseminating incivility in the Brazilian Twitter/X sphere. Third,
individual political influencers not only disseminate uncivil information by retweeting politicians, media agents, and other individual accounts but also generate a large amount of original uncivil texts. We find that the main mechanism of incivility delivery is a direct flow from political influencers to their followers. However, as influencers tend to retweet other influencers with shared followers, a mix of direct and indirect information delivery, rather than the traditional two-step flow of information, is more prevalent than expected at random.\\

While previous studies suggest that extremist right-wing groups predominantly instigate offline political violence, such as the attack of the U.S. Capitol after the 2020 election and of the Brazilian Congress after the 2023 election, the evidence from Brazilian Twitter/X space reveals a different dynamic: left-aligned individuals emerge as the primary drivers of uncivil information. On close inspection, many left-aligned users appear to be primarily responding to the uncivil behavior of their outgroup. This finding has two key implications. First, it underscores the existence of a feedback loop that exacerbates polarized tensions between political groups. Second, it highlights the efforts of online users to draw public attention to acts of violence. Future research should replicate this study across different time periods, countries, and platforms, and further differentiate direct and indirect mentions of incivility.

\section{Related Work}
\subsection{Online Incivility as a Multidimensional Concept }

\begin{figure}[t]
\centering
\includegraphics[width=1\linewidth]{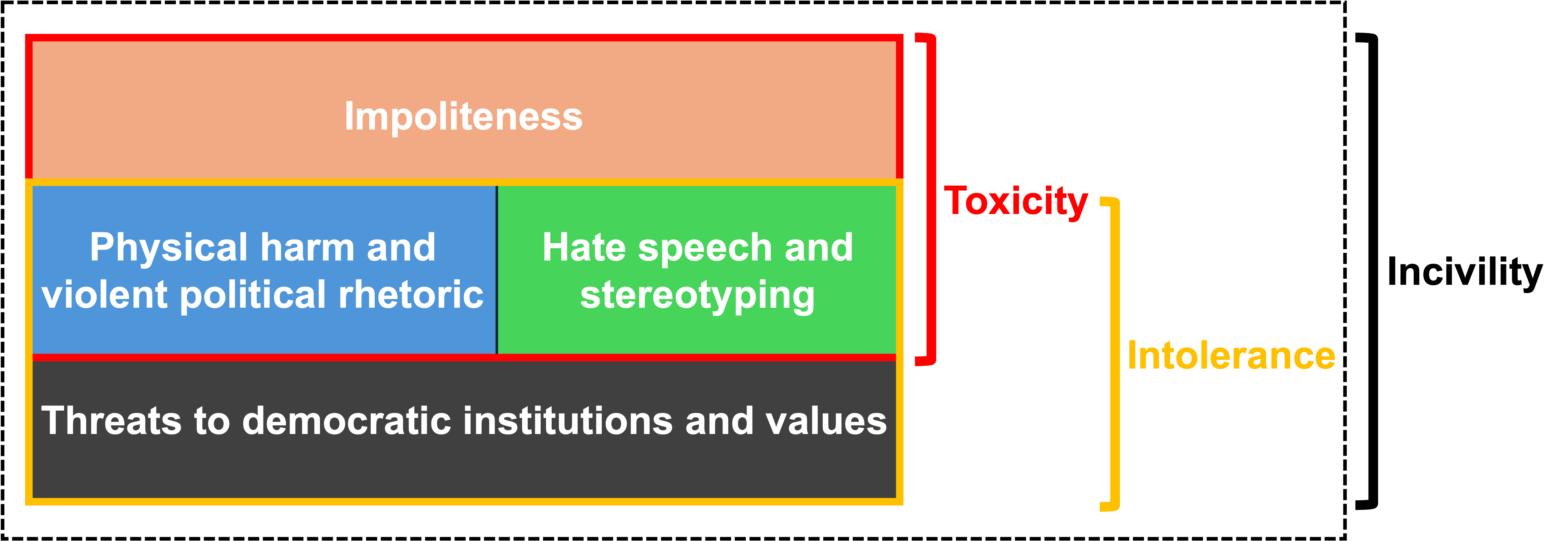}
\caption{Multi-dimensional conceptualization of incivility and its related concepts. Broadly speaking, incivility encompasses both impoliteness and intolerance, and toxicity overlaps with both concepts.}
\label{fig:conceptualization}
\end{figure}

Incivility broadly refers to behaviors that violate collective norms and hinder cooperation among social and political agents \cite{stryker2016political}. Related concepts include impoliteness, intolerance, and toxicity. Impoliteness focuses on disrespectful communication behaviors, while intolerance—referred to by different names in various literature and defined here following \citeauthor{rossini2022beyond}—highlights violations of democratic values in discourse \cite{papacharissi2004democracy, rossini2022beyond}. Toxicity overlaps with both concepts, emphasizing harmful elements directed at individuals or groups \cite{aleksandric2024users, chong2022understanding, singh2024differences}. To clarify these relationships, we provide a conceptual illustration in Fig.~\ref{fig:conceptualization}

The conceptualization of (political) incivility has long been a subject of debate. For example, \cite{gervais2015incivility} and \cite{coe2014online} focus on rude or disrespectful behaviors targeting individuals or groups, namely impoliteness. In contrast, \cite{papacharissi2004democracy} defines incivility as behaviors that undermine democratic values, aligning closely with intolerance as described by \cite{rossini2022beyond}. 

Recent literature has integrated these perspectives, proposing multidimensional frameworks for a more nuanced understanding of incivility \cite{gao2024crisis, muddiman2017personal, bentivegna2022searching}. For example, \cite{muddiman2017personal} distinguishes between personal-level and public-level incivility, while \cite{bentivegna2022searching} categorizes incivility into dimensions such as impoliteness, individual delegitimization, and institutional delegitimization. Our approach summarizes and provides conceptual clarity to these various works. It stems from the collective norms that incivility violates. \footnote{The incivility dimensions in this work are part of the Swiss National Science Foundation (SNSF) project: \textit{From Uncivil Disagreement to Political Unrest? A Cross-Platform \& Cross-National Analysis of the Offline Consequences of Online Incivility.}}

\subsection{Quantifying Online Incivility}
The widespread availability of large-scale data and recent advances in machine-learning techniques have enabled the automatic detection of online incivility. Among the most commonly measured categories are toxicity and hate speech \cite{gitari2015lexicon, badjatiya2017deep, shah2021deep}. For instance, Google and Jigsaw developed the Perspective API, which uses machine learning to score the toxicity of input text. Similarly, hate speech, defined as "abusive speech targeting specific group characteristics, such as ethnicity, religion, or gender", has been tackled using various automated methods, including TF-IDF \cite{akuma2022comparing}, lexicon-based approaches \cite{gitari2015lexicon}, deep learning methods \cite{badjatiya2017deep}, and hybrid approaches \cite{shah2021deep}.

Hate speech has also been distinguished from other impolite categories, such as offensive, abusive, aggressive, and cyberbullying language \cite{founta2018large}. At a more granular level, hate speech has been subcategorized into types such as religion-based, sexist, racist, homophobic, and other forms of hate \cite{gomez2020exploring}. In computer science, most models for detecting hate speech are evaluated on generalized datasets \cite{davidson2017hate, mathew2021hatexplain, mollas2022ethos}. However, researchers in other domains often fine-tune models for downstream applications for specific tasks and develop custom classifiers tailored to their unique requirements e.g., \cite{theocharis2020dynamics, frimer2023incivility}.

To predict uncivil categories, previous studies have utilized logistic regression models \cite{theocharis2020dynamics}, neural networks \cite{maity2018opinion}, and more recently, pre-trained language models such as BERT and RoBERTa \cite{davidson2020classifier, gao2024crisis}. For shorter social media texts, \cite{reimers2019sentence} introduced sentence transformers, which pool word embeddings into sentence-level representations to capture contextual meaning more effectively. However, even though classifiers for identifying toxicity and hate speech are common, classifiers targeting violations of democratic values, are rare. To the best of our knowledge, our work develops the most comprehensive classifiers, covering various dimensions of incivility. 

\subsection{Online Political Influencers and Multi-step Flow of Uncivil Information}

As social media increasingly dominates communication, we have entered an era where "any person can theoretically build an audience and grow their influence" \cite{goodwin2023political}. Online influencers are particularly effective in branding and marketing due to their authentic, celebrity-like connections with broad audiences. This study focuses on political influencers, defined as influential accounts that are dedicated to promoting political stances, social identities, or preferred candidates through their content \cite{riedl2023political}.

Prior research identifies three key criteria for classifying political influencers: (1) their content must be political, either aggregating general political information \cite{su2024role} or reflecting specific political or social identities \cite{zhang2023network}; (2) their influence can be reflected by proxies such as follower counts, diffusion cascades, and potential post-earnings \cite{bakshy2011everyone}; and (3) they leverage platform-specific features to promote content within distinct social media ecosystems \cite{harris2023dont}. Here, we identify political influencers from keywords self-disclosed in their profiles and their follower counts.

Political influencers broadly include politicians, media outlets, journalists, and any influential individuals \cite{mcgregor2020taking, marozzo2018analyzing, lee2015double}. Among these, individual influencers play an increasingly significant role in disseminating online information \cite{wu2011whosays}. These individual influencers are usually partisans, candidate supporters, and members of marginalized groups such as religious minorities, women, LGBTQ individuals, and ethnic communities.

The two-step flow of communication theory highlights the role of individual opinion leaders as intermediaries, relaying information from mass media to the broader audience \cite{katz1955interpersonal, katz1957two}. Extensions of this model—such as horizontal two-step flows, where opinion leaders also generate their own content, and multi-step or mixed-flow frameworks, where information travels through various channels—help capture the greater complexity of modern communication. \cite{hunt2024horizontal, hilbert2017onestep}. Therefore, this work tests not only the two-step information flow and information from media sources, but also direct and mixed flows originating from different influential actors. Social media platforms now allow researchers to approximate these flows using platform-specific interactions. For instance, on Twitter/X, following relationships often represent direct information flows, while retweets may reflect two-step flow processes \cite{hemsley2019followers}.

\section{Data and Methods}
\subsection{Data Collection and Socio-political Identity Annotation}
During the 2022 Brazilian presidential election—held on October 2 (first round) and October 30 (runoff), NetQuest, an international survey company, conducted a national survey. The survey involves 1,018 respondents who are representatively sampled by gender and region, with targeted income distribution nationwide. Data were collected through Netquest’s proprietary online panel using stratified quota sampling to ensure demographic representativeness. Panelists were recruited via double opt-in procedures and completed the survey online in exchange for incentives. Respondents were asked for consent to provide their Twitter handles. Of the 1,018 respondents, 403 consented, and 271 were verified as existing Twitter accounts, accounting for 26.62\% of the total panel.

We collect the Twitter accounts followed by the 271 survey participants using the Twitter API, resulting in 57,645 followers and 73,755 following pairs. We then identify political influencers based on three criteria: 1) having at least 1,000 followers, 2) profile description displaying politics-related content \cite{khamis2017self, harff2023influencers}, and 3) being located in Brazil (see Section 2 in Appendix for more details).

Ultimately, we obtain data from 204 survey participants, 2,307 political influencers, and 4,107 following pairs. Additionally, we collect tweets posted or retweeted by political influencers from 1st September 2022 to 1st February 2023, resulting in 5.22 million tweets. 

To evaluate the representativeness of the 204 survey sample, we compare several demographic variables, including Age, Gender, Ethnic, Religion, Income, and Education, between the 204 sample and the full survey sample. Pearson's chi-squared tests (for categorical variables) and Mann–Whitney U tests (for discrete variables) of the demographic variables between the 204 sample and the 1,018 respondents do not reject the null hypothesis that both samples are drawn from the same distribution at a significance level of 5\% (see Fig.~\ref{fig:sample validation} in Appendix). Fig.~\ref{fig:ideology distribution} in Appendix further demonstrates that the ideological position of the 204 sample has no strong bias toward any particular leaning. We then examine how this representative survey sample is exposed to incivility disseminated by the political influencers they follow.

We manually annotate the socio-political identities of uncivil political influencers. A human coder who is familiar with Brazilian politics first classifies each influencer by Account Type—categorizing them as politicians, media (including media outlets and journalists), or individual influencers. The coder further annotates their socio-political identities based on three dimensions: Ideological Position (Left/Right/Center), Campaign Support (Lula/Bolsonaro), and Social Identity (including Women—particularly in the context of advocating for women's rights, Religious, Black, or LGBTQ). The choice of social identity categories is informed by a manual review of 2,000 randomly sampled profiles. A codebook detailing how these categories are identified is provided in Section 3 of the Appendix. These annotations are based on self-disclosed information from users’ Twitter/X profiles.  

\subsection{Multidimension Construction and Automatic Classification of Uncivil Texts}

We address RQ1 by using incivility dimensions that represent four collective norm violations, as predefined in our codebook (see Section 4 in Appendix). First, impoliteness is a typical dimension representing the violation of etiquette rules \cite{rega2023incivility}. Physical harm and violent political rhetoric extend beyond the realm of etiquette and signify violations of general norm of non-violence \cite{miller1984use}. Additionally, we consider behaviors that breach social equality \& non-discrimination, and democratic norms.  Hate speech and stereotyping are typical forms of uncivil language that violate social equality and non-discrimination, while violations of democratic norms involve political behaviors discrediting democratic institutions or values. 

We develop four supervised learning classifiers to identify multidimensional incivility, trained on a Brazilian corpus collected from multiple social media platforms during the 2022 Brazilian presidential election, including Twitter/X, Facebook, YouTube, Telegram, etc. The full list of data sources is provided in Section 4 of the Appendix. 

We select samples from this corpus for manual annotation by two human coders who are familiar with Brazilian politics and fluent in the Brazilian language. To ensure consistency, we conduct a pilot coding procedure involving the two coders and researchers. First, approximately 30 samples for each incivility dimension are annotated independently by the two coders. Second, a meeting is held between the researchers and the coders to discuss the samples with disagreements. Third, a final decision is made on the coding guidelines, and the process is repeated until the majority of samples are consistently agreed upon by both coders. Through five rounds of pilot annotations, we refine definitions to achieve consensus and clarity. Finally, each coder is assigned with 500 samples per dimension for formal annotation. There are 100 overlapping samples between the two coders' annotation sets used to assess inter-coder reliability. Upon agreement, we obtain a total of 900 annotated samples per dimension that serve as the initial training data for the supervised classifiers.

\begin{figure}[t]
\centering
\includegraphics[width=1.0\linewidth]{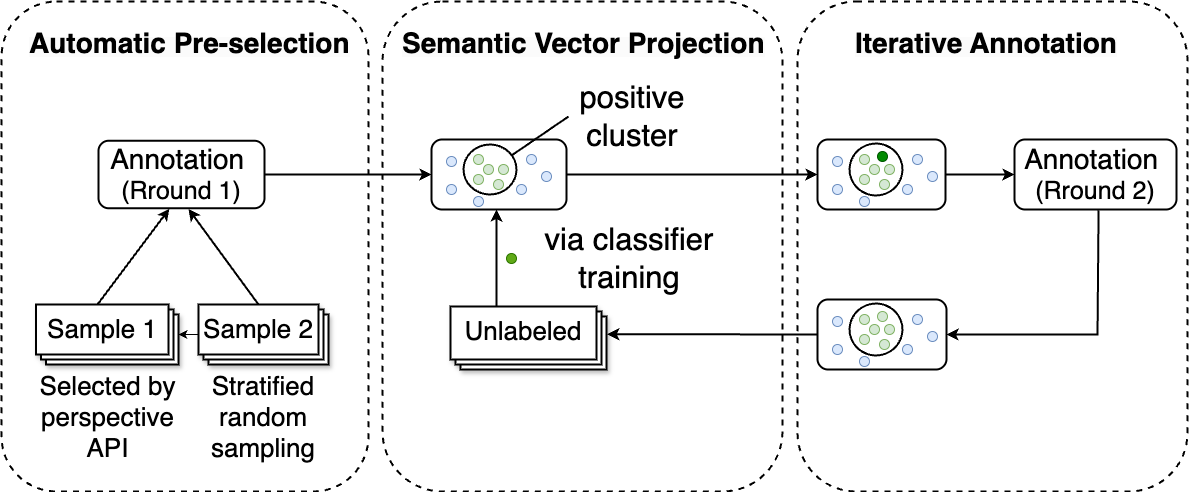}
\caption{Procedure for creating the positive (uncivil) samples with active learning.}
\label{fig:training}
\end{figure}

The initial annotation shows that positive (uncivil) samples of four dimensions of incivility accounts for 26.66\%, 1.52\%, 6.29\%, 6.30\% of the overall datasests repectively. To address the scarcity of positive samples, we employ an active learning strategy combining human annotation and automated modeling (see Fig.~\ref{fig:training}). Here is a detailed description of the three steps of this process:

\textbf{Step 1: Automatic Pre-selection.}  
We create an initial training dataset by combining posts flagged as toxic by the Perspective API with stratified random samples. Two coders who are familiar with Brazilian politics and Brazilian language annotate this dataset.

\textbf{Step 2: Semantic Vector Projection.}  
Annotated posts are mapped into a semantic vector space using pre-trained SentenceTransformer embeddings \cite{reimers2019sentence}. The centroid of the positive class is calculated as the mean of its document-level vectors. Unlabeled posts are compared to this centroid using cosine similarity to identify potential positive samples.

\textbf{Step 3: Iterative Annotation.}  
Unlabeled posts with high similarity to the positive centroid are sampled for manual annotation in a second round. To avoid potential bias, we also include posts with lower similarity, but not too far from the positive centroid. Concurrently, an initial classifier is trained on the limited data and used to filter unlabeled posts for additional positive examples. Annotated posts with agreement between coders are added to the training data. This approach enhances the diversity of positive cases.

\subsection{Investigating Dynamics of Mutil-dimensional Incivility Occurrence}

This section investigates the trends and events dynamics of incivility at various stages of the election, as questioned in RQ2. The trends refer to the upward and downward changes over time, while the events-level dynamics focus on outlining spikes on specific dates, often corresponding to the emergence of particular events. We utilize smoothing spline, a non-parametric regression method, to fit the trends in the frequency of uncivil posts and examine the outliers across various dimensions of incivility.

The loss function of the smoothing spline is given by $\sum_{i=1}^n \left(y_i - f(x_i)\right)^2 + \lambda \int f''(x)^2 \, dx,$ where \( f(x) \) represents the smoothing spline function and $n$ is the number of observations. The left-hand term measures the goodness-of-fit, and the right-hand term is the roughness penalty. The smoothing parameter \( \lambda \) controls the trade-off between the two terms.
First, we set the parameter \( \lambda \) to 0.6 to examine the trending patterns. We choose this value based on manual testing to ensure sufficient smoothness for capturing the trends. Second, to detect outliers representing unexpected events, we choose \( \lambda \) based on generalized cross-validation (GCV), which minimizes the predictive error \cite{wahba1990spline}.

We then identify the outliers that most significantly influence the model fitting using approximate Cook's distance, which quantifies the influence of individual data points on the estimated regression coefficients \cite{wood2017generalized}. Unlike the trends of incivility's general ups and downs, outliers indicate specific dates when uncivil content exhibits substantial spikes.

\subsection{Investigating Disseminators and Audience of Mutil-dimensional Incivility}

To address RQ3, we analyze the disseminators of incivility and their audiences (survey users)  across multiple dimensions. We first examine the number of survey users exposed to incivility at varying density levels. Incivility density is defined as the ratio of uncivil posts to the total number of posts by influencers. The exposed users are counted as the sum of direct followers and indirect retweets receivers of uncivil messages. To assess how incivility density influences exposure counts, we apply a quantile regression model \cite{buchinsky1998quantile}. Unlike traditional linear regression, this model estimates conditional quantiles (e.g., median or other percentiles) of the response variable, making it well-suited for our data, given the wide range of exposed audience counts at each density level. Additionally, we calculate Jaccard similarity to measure the overlap of political influencers and the survey users exposed to them across quantiles and dimensions. 

We then examine the specific identities of political influencers who post different ratios of incivility. Influencers are categorized into four quantiles, ranging from low to high incivility density. Their identities are analyzed across quantiles and dimensions based on manual annotation of profile descriptions. We use the G-test to assess differences in identity group distributions across incivility density levels and dimensions.

\subsection{Investigating Mechanisms of Information Flow of Mutil-dimensional Incivility}

Finally, we investigate how uncivil information is spread between political influencers and the survey audience (RQ4). To analyze the dissemination mechanisms of uncivil information, we construct a bipartite network \( G = (U, V, E) \), where \( U \) represents the set of ordinary followers, \( V \) denotes the set of political influencers, and \( E \) comprises edges indicating follower-influencer relationships. We then project this bipartite network onto the influencer set \( V \), resulting in a unipartite network \( G' = (V, F) \). In \( G' \), an undirected edge \( e_{ij} \in F \) between influencers \( v_i \) and \( v_j \) exists if they share common followers, with the edge weight corresponding to the number of shared followers. Additionally, we incorporate another set of directed edges, $R$, containing all the retweets done by the political influencers in $V$. An edge $r_{ij}\in R$ from $v_i$ to $ v_j$ represents retweets of $v_i$ by $v_j$, indicating the direction of the flow of information with a weight equal to the number of retweets. We also add a new set of nodes, $V'$, containing users who are retweeted by political influencers but are not in $V$, resulting in the retweet graph $G''= (V\cup V', R)$. 
The resulting network is a multilayer graph, $H$, with one layer, $G'$, representing shared followers of political influencers and a second layer, $G''$, capturing retweeting among them. Using the follower graph, $G$, and the multilayer graph $H$, we capture how survey users are exposed to uncivil content. This can happen either by being exposed to content created by an influencer they follow or when an influencer they follow retweets someone else. More precisely, we distinguish  three distinct motifs of information flow from political influencers to survey respondents: 

\begin{enumerate}
    \item \textbf{Direct flow}: direct exposure to an original tweet or a self-retweet by a political influencer followed by a survey user. 
    \item \textbf{Two-step flow}: indirect exposure occurs when a political influencer followed by a survey user retweets content from an account the survey user does not follow.
    \item \textbf{Mixed flow}: a mix of direct and indirect exposure when a survey user follows a political influencer who retweets another account also followed by the survey user.
\end{enumerate}

We note that the assumption above—that ordinary users are exposed to uncivil content solely through direct followership and indirect retweets—is a simplified proxy. In reality, users may also encounter such content through random browsing, private messages, or quotations that are not captured by our methods, meaning their actual exposure may differ from our estimates. Moreover, according to the traditional two-step information flow theory, paraphrased content from opinion leaders may also be relevant, which requires more sophisticated strategies.

\begin{figure}[t]
\centering
\includegraphics[width=1.0\linewidth]{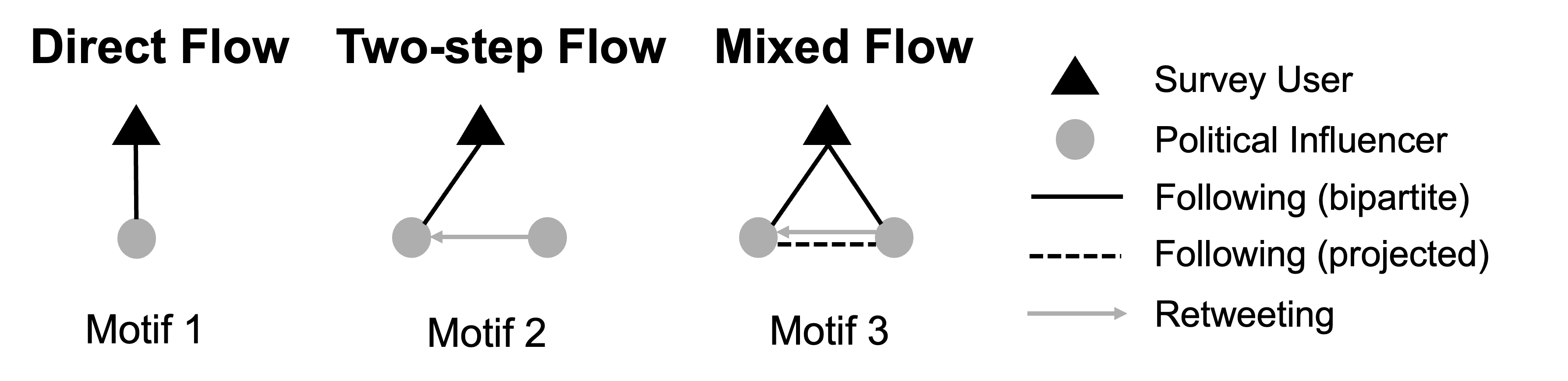}
\caption{Motifs of information flow mechanisms.}
\label{fig:information flow}
\end{figure}

We show the three motifs in Fig. \ref{fig:information flow}. To identify salient mechanisms, we calculate the $Z$-scores by comparing the observed motif counts with those obtained by randomizing the retweet edges using a directed configuration model. To determine who utilizes these mechanisms, we also analyze the identities of influencers within each motif and identify the leading creators of incivility using PageRank centrality.

\section{Results}

\subsection{Performance of Automatic Classification of Uncivil Texts}
 
 The related collective norms, dimensions, definitions, examples, and inter-coder reliability metrics of our automatic classification of uncivil texts  are summarized in Tab. \ref{tab:incivility taxonomy}.

\begin{table*}[htbp]
\centering
\renewcommand{\arraystretch}{1.2} 
\resizebox{\textwidth}{!}{ 
\begin{tabular}{|>{\raggedright\arraybackslash}p{2.6cm}|>{\raggedright\arraybackslash}p{3.2cm}|>{\raggedright\arraybackslash}p{6.5cm}|>{\raggedright\arraybackslash}p{4.5cm}|>{\raggedright\arraybackslash}p{2.8cm}|}
\hline
\rowcolor[HTML]{EFEFEF} 
\textbf{Related Norm} & \textbf{Dimension} & \textbf{Definition} & \textbf{Example (translated)} & \textbf{Gwet's AC2 Score/Agreement} \\ \hline

Etiquette & Impoliteness (IMP) & Messages displaying rudeness or disrespect, often using offensive language directed at individuals or groups. & \# Haven't you gotten tired of the "biggest corruption scheme in history" yet? Litany of idiots. & 0.78/0.85 \\ \hline

Non-violence & Physical Harm and Violent Political Rhetoric (PHAVPR) & Messages threatening physical harm against individuals or groups, or promoting violence in political contexts. & \# The Northeast also fights against the left! & 0.93/0.93 \\ \hline

Equality and non-discrimination & Hate Speech and Stereotyping (HSST) & Messages promoting hostility or discrimination against individuals or groups based on specific attributes, often referring to social identity groups. & \# Bolsonaro supporters are stupid in terms of numbers, gender and degree. & 0.90/0.91 \\ \hline

Democratic norms & Threats to Democratic Institutions and Values (THREAT) & Messages undermining democratic procedures and institutions, the democratic state, and democratic values. & \# The judiciary persecutes those who do not follow their rules...Our only hope is \#SOSFFAA Salve O Brasil & 0.92/0.92 \\ \hline

\end{tabular}}
\caption{Summary of related norms, dimensions, definitions, examples, and inter-coder reliability statistics of incivility. Examples are selected from our Brazilian Twitter/X dataset. Both Gwet's AC2 score and agreement are reported to assess inter-coder reliability.}
\label{tab:incivility taxonomy}
\end{table*}

Model performance, evaluated using ten-fold cross-validation, is summarized in Tab.~\ref{tab:classification_results}. We calculate the metrics by aggregating all predictions from each fold to help mitigate issues caused by class imbalance within individual folds. The weighted F1 scores across dimensions range from 78\% to 93\%, indicating good classification performance. The trained classifiers are subsequently applied to the remaining tweets of political influencers. To ensure reliable labeling, uncivil labels - "1" are assigned only when the predicted probabilities for the positive class are at least 0.7. The number of classified uncivil posts for each dimension is shown in Tab.~\ref{tab:summary-statistics}.

\begin{table}[htbp]
\renewcommand{\arraystretch}{1.2} 
    \centering
    \begin{tabularx}{\columnwidth}{@{}l>{\centering\arraybackslash}p{1cm}>{\centering\arraybackslash}p{0.9cm}>{\centering\arraybackslash}p{1cm}>{\centering\arraybackslash}p{1.9cm}@{}}
        \toprule
        \textbf{Classifier Type} & \textbf{Dim} & \textbf{F1\textsubscript{Dim}} & \textbf{F1\textsubscript{non-Dim}} & \textbf{F1\textsubscript{weighted}} \\ 
        \midrule
        Single & $D_\text{IMP}$ & 0.93 & 0.94 & 0.93 \\ 
        Ensemble & $D_\text{PHAVPR}$& 0.65 & 0.91 & 0.86 \\ 
        Ensemble & $D_\text{HSST}$ & 0.76 & 0.79 & 0.78 \\ 
        Ensemble & $D_\text{THREAT}$ & 0.75 & 0.84 & 0.80 \\ 
        \bottomrule
    \end{tabularx}
    \caption{Cross-validation results of classification performance for the four uncivil dimensions.}
    \label{tab:classification_results}
\end{table}

\begin{table}[htbp]
\renewcommand{\arraystretch}{1.2} 
    \centering
    \begin{tabularx}{\columnwidth}{@{}l p{1cm} p{1cm} p{1cm} p{1cm} @{}}
        \toprule
        & \textbf{\boldmath$D_\text{IMP}$} & \textbf{\boldmath$D_\text{PHAVPR}$} & \textbf{\boldmath$D_\text{HSST}$} & \textbf{\boldmath$D_\text{THREAT}$} \\ 
        \midrule
        N (Posts) & 138,153 & 34,283 & 46,522 & 107,695 \\ 
        N (Influencers) & 1,688 & 1,598 & 1,481 & 1,631 \\ 
        N (Followers) & 199 & 200 & 198 & 200 \\ 
        \bottomrule
    \end{tabularx}
    \caption{Summary statistics of the classified dataset.}
    \label{tab:summary-statistics}
\end{table}

\subsection{Different Dimensions of Incivility Break Out at Various Stages}

\begin{figure*}[t]
\centering
\includegraphics[width=1.0\linewidth]{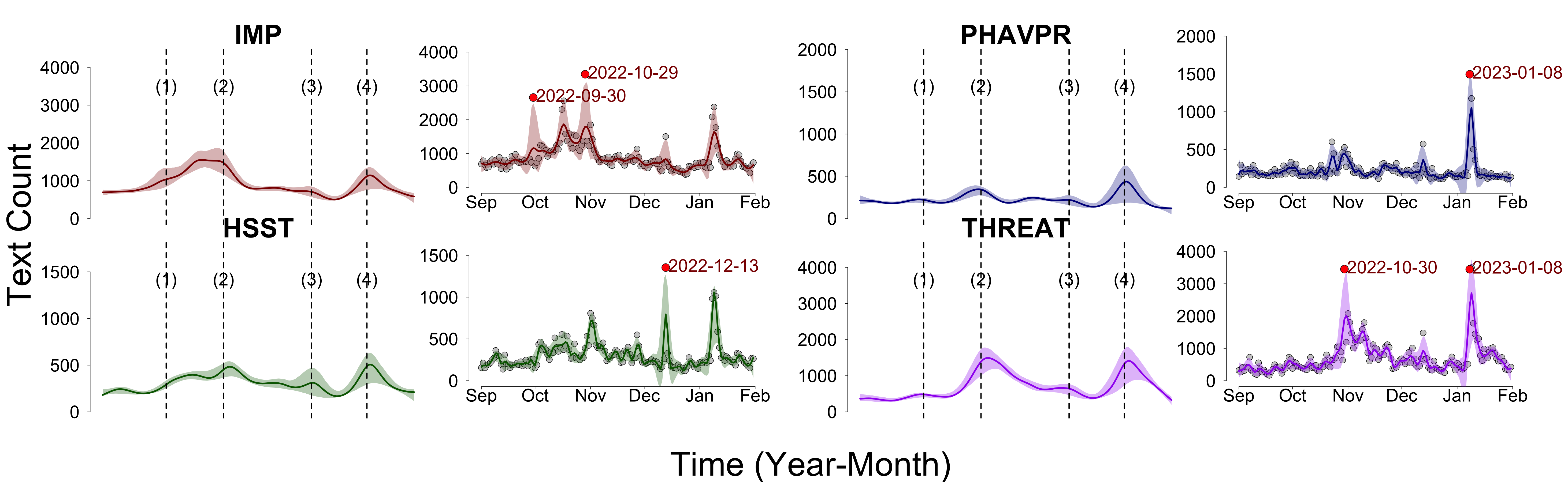}
\caption{Trends and events dynamics of four incivility dimensions—IMP, PHAVPR, HSST, and THREAT. The first and third columns show trends with smoothing splines ($\lambda$=0.6). Original data points with GCV smoothing splines and outliers (red dots) identified by approximate Cook's distance are in the second and fourth columns. Grey lines mark five key events: (1) First-round election, (2) Run-off election, (2) Brasília protest, (3) Highway blockades, and (4) Congress attack.}
\label{fig:dynamics}
\end{figure*}

Here we examine when during the Brazilian election campaign different dimensions of incivility appear (RQ2).
The fitting lines in Fig.~\ref{fig:dynamics} illustrate the trends and key events for different dimensions of incivility during the election period. IMP shows a significant upward trend around the two election rounds and the Congress attack, with outliers highlighting key dates just before the first-round election (October 2, 2022) and the runoff election (October 30, 2022). PHAVPR exhibits pronounced activity during the Congress attack, with outliers pinpointing January 7 and 8, 2023, as significant dates. HSST, besides peaking during the two election rounds and the Congress attack, also shows a notable increase during the road and highway blockades, with December 13, 2022, identified as a key date. THREAT displays a macro trend with two major peaks: the run-off election (coinciding with the Brasília protest) and the Congress attack, with outliers on December 30, 2022, and January 8, 2023.

In summary, all dimensions of incivility surge around key events, but their magnitudes vary over time. IMP is most concentrated during the election rounds, suggesting its use in voter attraction and election-related strategies. In contrast, PHAVPR, HSST, and THREAT are more prevalent during violent offline events happening, reflecting their association with violence and unrest. Outlier detection further validates these patterns, highlighting distinct temporal dynamics across dimensions of incivility.

\subsection{Dominant Disseminators Differ but Share Similar Identities Across Dimensions}

\begin{figure}[t]
\centering
\includegraphics[width=1.0\linewidth]{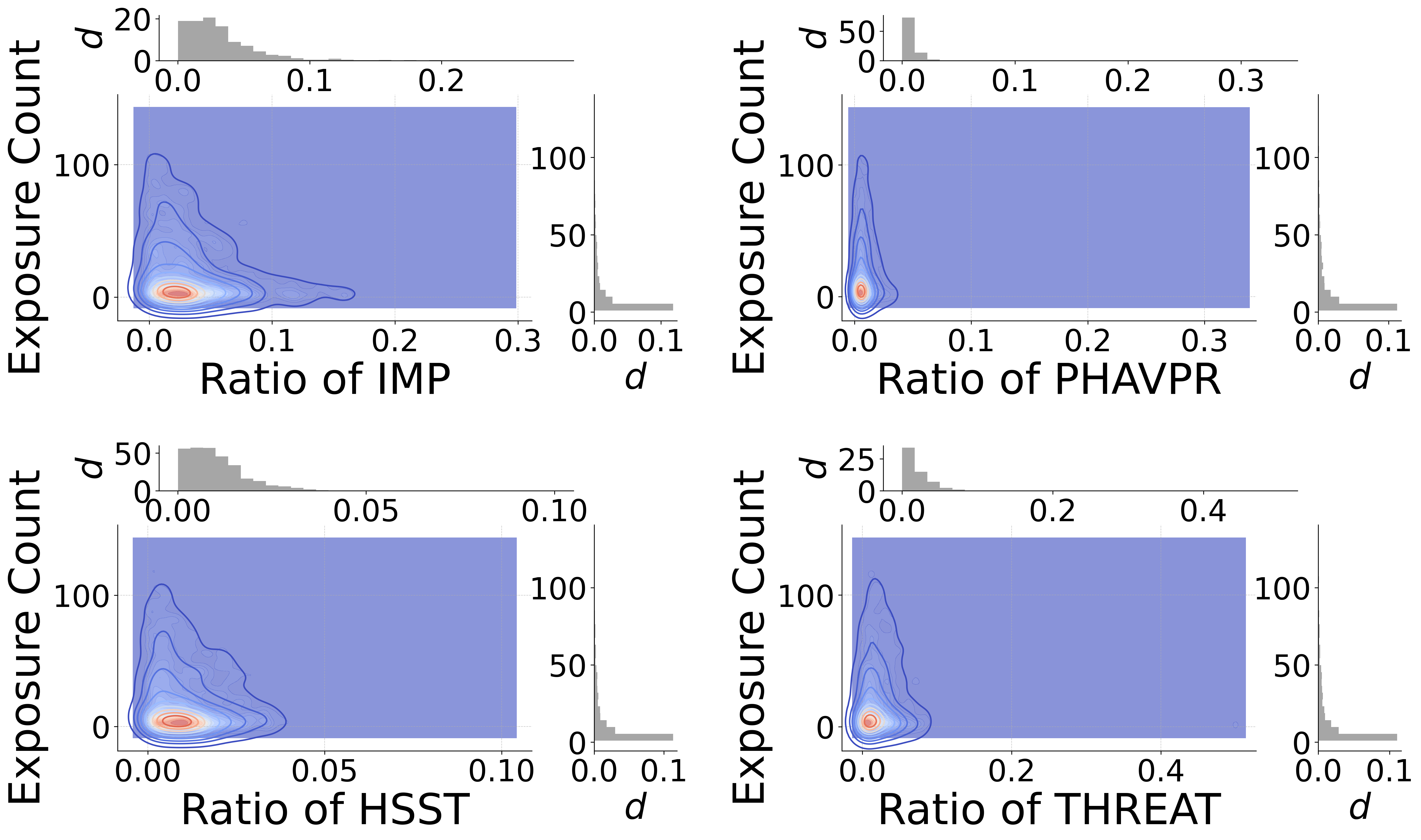}
\caption{Kernel density estimate plot of the variation in survey audience's exposure, measured by the total number of followers and retweet receivers of uncivil messages, as the ratio of uncivil posts (density) increases.}
\label{fig:influencer and audience distribution}
\end{figure}

This section identifies the main disseminators (political influencers) and audiences (survey users) of online uncivil messages (RQ3). First, disseminators are ranked by their ratio of uncivil posts to total posts, from low to high. Fig.~\ref{fig:influencer and audience distribution} presents the distributions of exposed audiences, calculated as the sum of followers and retweet receivers of uncivil disseminators, across different levels of uncivil ratios. Quantile regression results (Tab.~\ref{tab:quantile regression}) reveal that for all dimensions except THREAT, the number of exposed audiences decreases as disseminators' uncivil ratios increase. The correlation is strongest for IMP between the 0.25 and 0.9 quantiles, reaching a 5\% significance level. For PHAVPR and HSST, significance is observed only at the median level. Conversely, for THREAT, the number of exposed audiences significantly increases at the 0.9 quantile as the incivility ratio rises. This suggests that more frequent uncivil posts do not necessarily attract larger audiences.

Disseminators and their audiences are further divided into four quantile subgroups based on the disseminators' uncivil ratio rankings for each incivility dimension. The average value of Jaccard similarity of disseminators across dimensions and quantiles demonstrates high dissimilarity ($0.1035 \pm 0.0707$). However, their audiences exhibit substantial overlap across dimensions and quantiles ($0.8827 \pm 0.0460$). This raises important questions about the identities of the main disseminators across dimensions and levels and why they attract similar audiences.

\begin{table*}[htbp]
    \centering
    \begin{tabularx}{\linewidth}{l*{5}{>{\centering\arraybackslash}X}}
        \toprule
        \textbf{DIM} & \textbf{Q(0.1)} & \textbf{Q(0.25)} & \textbf{Q(0.5)} & \textbf{Q(0.75)} & \textbf{Q(0.9)} \\ 
        \midrule
        $D_\text{IMP}$ & $-1.45 \cdot 10^{-6}$ & $-7.58^{**}$ & $-29.51^{***}$ & $-111.89^{***}$ & $-250.97^{***}$ \\ 
        $D_\text{PHAVPR}$ & $-6.98 \cdot 10^{-6}$ & $-17.60^{*}$ & $-56.54^{***}$ & $-149.34$ & $-136.11$ \\ 
        $D_\text{HSST}$ & $-3.78 \cdot 10^{-6}$ & $-11.04$ & $-63.61^{**}$ & $-263.59$ & $-550.13^{*}$ \\ 
        $D_\text{THREAT}$ & $-1.06 \cdot 10^{-6}$ & $-2.03$ & $-11.36$ & $-2.90 \cdot 10^{-7}$ & $222.90^{***}$ \\ 
        \bottomrule
    \end{tabularx}
    \caption{Quantile regression results for exposure and incivility ratio across dimensions}
    \label{tab:quantile regression}
\end{table*}

\begin{figure}[t]
\centering
\includegraphics[width=0.9\linewidth]{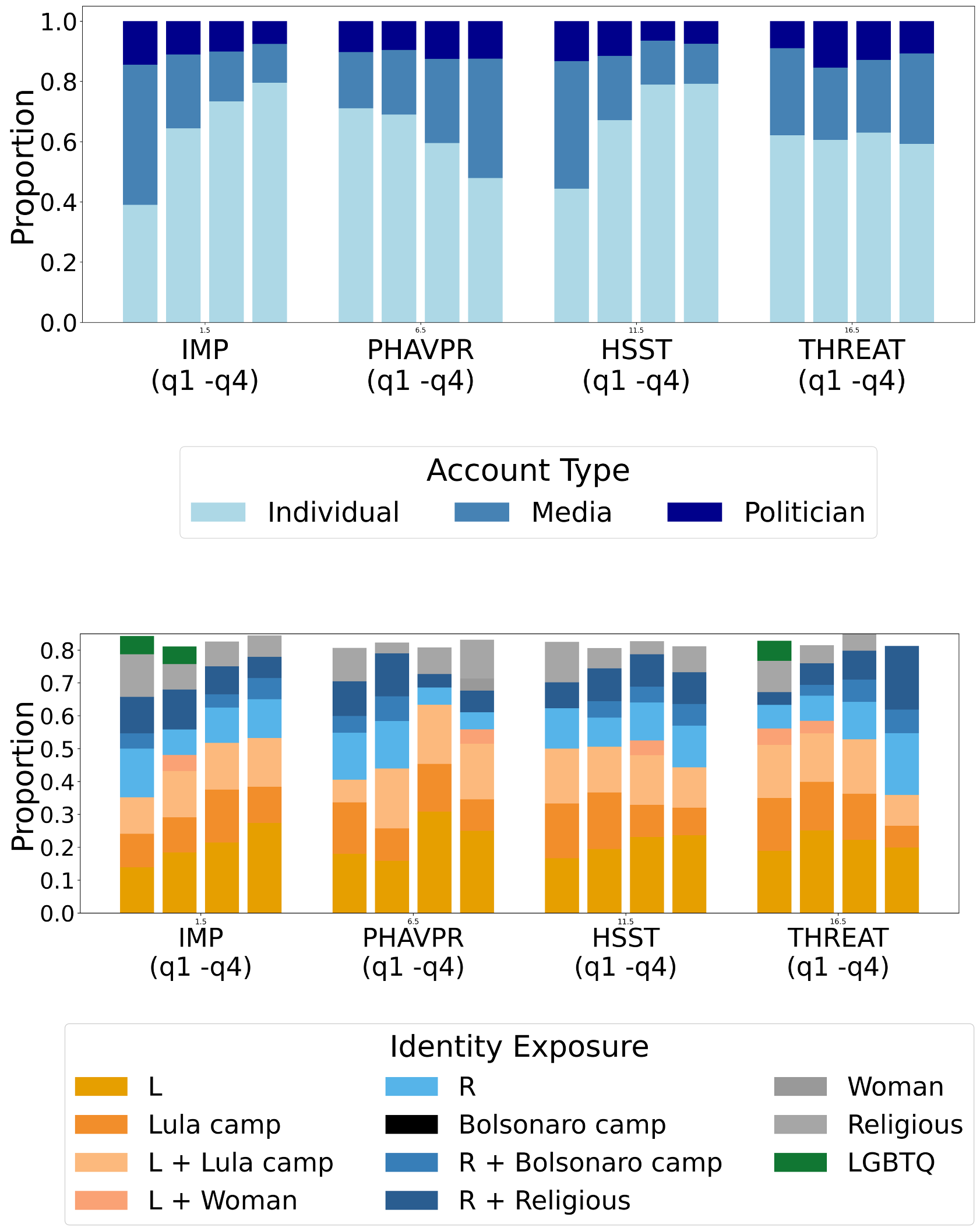}
\caption{Distribution of account types and individual self-disclosed identities across quantiles of uncivil densities ($q_1, q_2, q_3, q_4$, from low to high) and dimensions (IMP, PHAVPR, HSST, THREAT). Categories are ranked in descending order, and only those that cumulatively account for 80\% of all individual identities are displayed.}
\label{fig:identity distribution}
\end{figure}

We show the distributions of disseminator account types across incivility dimensions and quantiles of incivility ratios in Fig.~\ref{fig:identity distribution}. Accounts from individuals constitute the majority of political influencers across all dimensions: IMP (63.08\%), PHAVPR (60.59\%), HSST (66.39\%), and THREAT (59.18\%). For IMP and HSST, the proportion of individual accounts is larger, while the proportions of politicians and media accounts is smaller for higher levels of incivility ratio. Conversely, for PHAVPR, the proportion of individual accounts becomes smaller and the proportions of politicians and media agents larger for higher levels of incivility density. For THREAT, account type distributions remain relatively stable. The G-test statistics (Tab.~\ref{tab:g-test}) confirm significant differences in account type distributions at the 5\% significance level for IMP, PHAVPR, and HSST, but not for THREAT.

The distribution of self-reported identities among individual accounts shows that left-aligned individuals, such as those with left-wing ideologies or supporting Lula, constitute a larger proportion of individual influencers across all dimensions: IMP (43.12\%), PHAVPR (44.42\%), HSST (43.54\%), and THREAT (44.30\%). In comparison, right-aligned individuals, such as those with right-wing ideologies or supporting Bolsonaro, account for smaller proportions: IMP (22.91\%), PHAVPR (22.74\%), HSST (24.06\%), and THREAT (24.26\%).

For IMP, the proportion of left-aligned individuals increases with incivility level. For PHAVPR, while there is no strictly ascending trend across quantiles, left-aligned individuals are significantly more prevalent in the third and fourth quantiles than in the first and second. In contrast, for THREAT, the proportion of left-aligned individuals decreases as incivility levels rise, while HSST shows no significant changes. G-test statistics (Tab.~\ref{tab:g-test}) confirm significant differences at the 5\% level for all dimensions except HSST.

\begin{table}[t]
    \centering
    \arrayrulecolor{black} 
    \begin{tabularx}{\columnwidth}{@{}lXX@{}}
        \toprule
        \textbf{Dimension (DIM)} & \textbf{G-test Chi$^2$ (AT)} & \textbf{$p$-value (AT)} \\ 
        \midrule
        $D_\text{IMP}$       & $176.88$        & $1.56 \cdot 10^{-35}$ \\ 
        $D_\text{PHAVPR}$    & $59.54$         & $5.57 \cdot 10^{-11}$ \\ 
        $D_\text{HSST}$      & $130.58$        & $9.69 \cdot 10^{-26}$ \\ 
        $D_\text{THREAT}$    & $12.39$         & $5.38 \cdot 10^{-2}$ \\ 
        \midrule
        \textbf{Dimension (DIM)} & \textbf{G-test Chi$^2$ (IE)} & \textbf{$p$-value (IE)} \\ 
        \midrule
        $D_\text{IMP}$       & $115.02$        & $2.38 \cdot 10^{-2}$ \\ 
        $D_\text{PHAVPR}$    & $134.73$        & $3.72 \cdot 10^{-4}$ \\ 
        $D_\text{HSST}$      & $95.64$         & $1.81 \cdot 10^{-1}$ \\ 
        $D_\text{THREAT}$    & $148.10$        & $7.86 \cdot 10^{-6}$ \\ 
        \bottomrule
    \end{tabularx}
    \caption{Results of the G-test for account types (AT) and individual identities (IE) across the four quantiles of uncivil density for each dimension.}
    \label{tab:g-test}
\end{table}

\subsection{Direct Information Flow Dominates in Uncivil Communication Online, Whereas Mixed Information Flow is Significant}

To understand the mechanisms of information flow responsible for the spread of uncivil content (RQ4), we investigate the motifs in the network of retweets and co-follow relations. We distinguish the three information flow motifs—direct flow, two-step flow, and mixed flow (see Fig. \ref{fig:information flow}). The motif counts and $Z$-scores are presented in Tab. \ref{tab:motif-analysis}. The direct flow involves only one influencer, and the Z-score is therefore computed by only randomizing their self-retweets. 
Our results reveal that direct flow is the dominant pattern of incivility diffusion of uncivil content, two-step flow is the second most prevalent, and mixed flow is the least prevalent.
Interestingly, the $Z$-scores reveal that two-step flow motifs are significantly less prevalent, and mixed flow motifs are significantly more prevalent than in networks where retweets have been randomized.
This shows that when political influencers retweet other accounts, they tend to favor accounts with whom they have shared followers. Therefore, survey users' tendency to be subjected to the same uncivil content in a mixed way is higher than expected at random.

\begin{table*}[htbp]
\renewcommand{\arraystretch}{1.2}
\centering

\begin{minipage}[t]{0.48\textwidth}
\centering
\textbf{Observed Motifs ($I$)}\\[1ex]
\begin{tabularx}{\textwidth}{lXXXX}
\toprule
\textbf{Motif} & $I_{\text{IMP}}$ & $I_{\text{PHAVPR}}$ & $I_{\text{HSST}}$ & $I_{\text{THREAT}}$ \\
\midrule
Direct    & 169700 & 90685 & 52473 & 197595 \\
Two-step  & 47051  & 19511 & 23059 & 95984  \\
Mixed     & 2983   & 909   & 1409  & 5278   \\
\bottomrule
\end{tabularx}
\end{minipage}
\hfill
\begin{minipage}[t]{0.48\textwidth}
\centering
\textbf{Z-scores ($Z$)}\\[1ex]
\begin{tabularx}{\textwidth}{lXXXX}
\toprule
\textbf{Motif} & $Z_{\text{IMP}}$ & $Z_{\text{PHAVPR}}$ & $Z_{\text{HSST}}$ & $Z_{\text{THREAT}}$ \\
\midrule
Direct    & 7.07   & 16.55 & -2.76 & -2.28 \\
Two-step  & -60.49 & -39.19 & -57.53 & -67.39 \\
Mixed     & 28.55  & 12.51 & 31.90 & 38.64 \\
\bottomrule
\end{tabularx}
\end{minipage}

\vspace{2em}

\begin{minipage}[t]{0.48\textwidth}
\centering
\textbf{Mean Randomized ($RM$)}\\[1ex]
\begin{tabularx}{\textwidth}{lXXXX}
\toprule
\textbf{Motif} & $RM_{\text{IMP}}$ & $RM_{\text{PHAVPR}}$ & $RM_{\text{HSST}}$ & $RM_{\text{THREAT}}$ \\
\midrule
Direct    & 169348.1 & 90206.4 & 52542.6 & 197759.1 \\
Two-step  & 50013.7  & 20666.8 & 24506.7 & 100740.6 \\
Mixed     & 1677.6   & 570.4   & 650.4   & 2817.7 \\
\bottomrule
\end{tabularx}
\end{minipage}
\hfill
\begin{minipage}[t]{0.48\textwidth}
\centering
\textbf{Std Dev Randomized ($RS$)}\\[1ex]
\begin{tabularx}{\textwidth}{lXXXX}
\toprule
\textbf{Motif} & $RS_{\text{IMP}}$ & $RS_{\text{PHAVPR}}$ & $RS_{\text{HSST}}$ & $RS_{\text{THREAT}}$ \\
\midrule
Direct    & 49.72 & 28.90 & 25.04 & 72.10 \\
Two-step  & 48.98 & 29.50 & 25.16 & 70.58 \\
Mixed     & 45.72 & 27.06 & 23.79 & 63.68 \\
\bottomrule
\end{tabularx}
\end{minipage}

\caption{The comparison of observed motifs with motifs in the randomized configuration model. Motifs 1--3 represent Direct Flow, Two-step Flow, and Mixed Flow. $I$ denotes the number of motifs in the observed network. $RM$ represents the average number of motifs in the randomized configuration model. $RS$ represents the standard error of motifs in the randomized configuration model. The difference between observed motifs and random motifs is expressed using $Z$-scores.}
\label{tab:motif-analysis}
\end{table*}

\begin{figure}[t]
\centering
\includegraphics[width=1.0\linewidth]{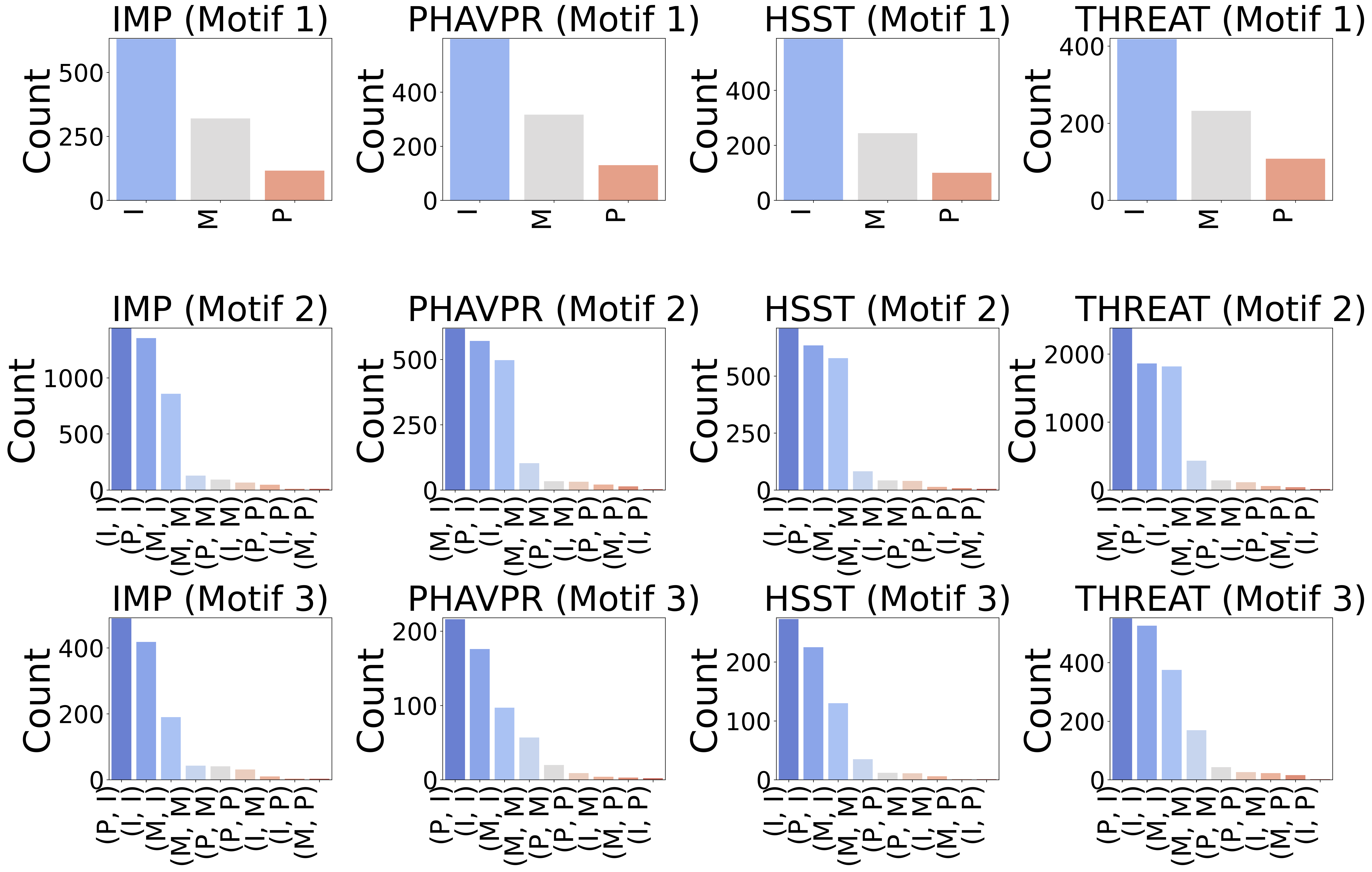}
\caption{Histograms displaying the distribution of account types—Individual (I), Media (M), and Politician (P)—of uncivil disseminators across different information flow motifs and dimensions of incivility. The label (A, B) indicates an information flow from A to B, namely B retweets A.}
\label{fig:identity in motifs}
\end{figure}

The account types of disseminators involved in these flow mechanisms are shown in Fig. \ref{fig:identity in motifs}. Across most dimensions and motifs, individual influencers play an important role in spreading uncivil information. They act both as intermediaries who retweet content from politicians, media, or other individual accounts (as shown in motif 2 and 3) and direct disseminators who create their own original uncivil posts (as shown in motif 1 and 3). 

Overall, the mixed information flow emerges as the most significant mechanism for disseminating uncivil content, as compared with the randomized retweet configuration model. Further analysis of the socio-political identities of individual influencers in Mixed Flow motifs shows that over 80\% of individuals retweeting uncivil messages from politicians, media, and other individuals are left-aligned users, such as those with left-wing ideologies or Lula supporters. This aligns with earlier findings highlighting the role of left-aligned individuals in transferring uncivil content from official political accounts to the public.

We also identify the dominant creators of incivility in retweet networks using PageRank centrality (see Tab.\ref{tab:politician_centrality}). Politicians are key creators of IMP messages, while media accounts, including outlets and journalists, dominate in producing THREAT messages. The creators of PHAVPR and HSST messages include a mix of politicians, media, and individual influencers. These results, alongside manual post-examinations, suggest that IMP is frequently used as a campaign strategy during elections, whereas dimensions like THREAT serve as denunciations of right-wing extremists.

\begin{table}[htbp]
    \centering
    \begin{threeparttable}
        \arrayrulecolor{black} 
        \begin{tabularx}{\columnwidth}{
@{}>{\centering\arraybackslash}p{0.2cm}
|>{\centering\arraybackslash}p{0.6cm}>{\centering\arraybackslash}p{0.4cm}
|>{\centering\arraybackslash}p{0.9cm}>{\centering\arraybackslash}p{0.4cm}
|>{\centering\arraybackslash}p{0.6cm}>{\centering\arraybackslash}p{0.4cm}
|>{\centering\arraybackslash}p{0.9cm}>{\centering\arraybackslash}p{0.4cm}
@{}}
            \toprule
            \textbf{\textit{\#}} & $D_\text{IMP}$ & \textit{L} & $D_\text{PHAVPR}$ & \textit{L} & $D_\text{HSST}$ & \textit{L} & $D_\text{THREAT}$ & \textit{L} \\ 
            \midrule
            1 & An** & PO & Ha** & PO & Ga** & PO & Gl** & MO \\ 
            2 & Lu** & PO & An** & PO & An** & PO & UO** & MO \\ 
            3 & La** & IN & An** & JL & Ja** & PO & Ca** & JL \\ 
            4 & Gl** & PO & Mo** & JL & Tr** & JL & Gl** & PO \\ 
            5 & Ca** & PO & Gi** & JL & NP** & IN & Re** & JL \\ 
            6 & Ha** & PO & Fo** & MO & Bo** & PO & Gu** & JL \\ 
            7 & De** & IN & Xi** & JL & La** & JL & Me** & MO \\ 
            8 & Jo** & PO & UO** & MO & Re** & JL & EC** & JL \\ 
            9 & Ja** & PO & Er** & IN & Se** & PO & Es** & MO \\ 
            10 & Fl** & PO & Th** & IN & Fo** & MO & Is** & IN \\ 
            \bottomrule
        \end{tabularx}
    \end{threeparttable}
    \caption{Rankings (\textit{\#}) of politicians based on PageRank centrality across four dimensions. Account types are explicitly labeled as PO (Politician), MO (Media outlet), JL (Journalist), IN (Individual), as shown in column L.}
    \label{tab:politician_centrality}
\end{table}

\section{Conclusion and Discussion}
This study automatically detects four dimensions of incivility—IMP, PHAVPR, HSST, and THREAT—on Twitter/X during the 2022 Brazilian Presidential Election and analyzes their dynamics, disseminators and audience, and dissemination mechanisms. The analysis of post dynamics reveals that the four dimensions of incivility emerge at different stages. IMP primarily peaks on election days during political campaign. In contrast, PHAVPR, HSST, and THREAT are more prominent while violent events happening. An analysis of the disseminators' identities indicates that left-aligned individuals are more prominently involved in spreading uncivil information. They demonstrate a significant pattern of retweeting uncivil information from politicians, media, and other influential accounts, and share common audiences with the accounts they retweet. This indicates that a mixed information flow, rather than pure two-step information flow, is the significant mechanism for spreading uncivil content. 

This finding contrasts with previous studies, which suggest that the right-wing exhibits more uncivil behavior during political campaigns e.g., \cite{rega2023incivility}. A manual post-check of the uncivil posts made by left-aligned and right-aligned influencers revealed that, except for IMP, a large number of uncivil posts are indirect mentions of incivility, meaning that users are discussing or even opposing incivility expressed by others. This phenomenon is more prominent among left-aligned influencers. For instance, "they are shooting at Indigenous people in the streets (PHAVPR)", "I express my full solidarity with an official who was targeted by a misogynistic and cowardly political supporter (HSST)", "We are vigilant and monitoring all necessary measures to curb coup-related and anti-democratic acts (THREAT)" \footnote{The examples provided are translated into English and paraphrased to protect user privacy.}. This could explain why PHAVPR, HSST, and THREAT are more prevalent during offline violent events led by right-wingers and are primarily associated with left-aligned political influencers. Moreover, determining whether violent metaphors imply actual harm is also challenging. Our results have to be interpreted in light of these limitations, and we advise future studies to exercise caution when applying the classifiers to downstream tasks involving such reverse discourse patterns and metaphors.

The indirect mentions of incivility, especially for dimensions PHAVPR, HSST, and THREAT, can also be found in the discourse of right-aligned political influencers. Even though this may not be called uncivil, indirect mentions of incivility still risk escalating into "violence for violence", where users respond to the uncivil behavior of their outgroup in similarly uncivil ways. For instance, we find some direct uncivil expressions from the left-aligned users, such as "I want to gather everyone from the inauguration and surround Brasília to see who’s stronger (PHAVPR)", "An end to these fascist Nazis disguised as Novo (HSST)", "It's time to confront the coup without mediation (THREAT)". Such cycles can be more detrimental to democratic processes than incivility itself. However, the indirect mentions of incivility can also be beneficial as they draw public attention to stop political violence and enable disadvantaged groups to be seen. Future research could further enhance automated techniques to distinguish between direct and indirect mentions of incivility and assess their respective impacts.

The recent global rise of right-wing populism has often been associated with increased political incivility and violence. In Brazil, this trend was reflected during the Bolsonaro administration (2018–2022). However, growing dissatisfaction with Bolsonaro’s pandemic response and perceived authoritarianism contributed to a left-leaning resurgence, culminating in Lula’s 2022 presidential victory. Lula’s campaign, driven by digital mobilization, has marked a renewed "pink tide" reminiscent of the early 2000s \cite{lampter2023pink}. This shift may account for the prevalence of both direct and indirect incivility among left-leaning users.

Additionally, the observed mixed information flow pattern suggests a strong tendency toward echo chamber formation on social media. Ordinary users who follow a political influencer are also more likely to follow another influencer who retweets that influencer. This dynamic is particularly pronounced among politicians, individual influencers, and their shared followers. Such patterns may be driven by platform algorithms and the increasing presence of individual supporters acting as political influencers. 

Like many previous studies, this study is not without limitations. For instance, the automatic classification models, while useful, are less accurate than human labeling—a limitation shared by other widely used uncivil detection models like Perspective API \cite{hosseini2017deceiving}. Additionally, we only collect data from political influencers associated with 204 survey users, a rather small sample. However, the sample's attribute distributions do not differ significantly from the overall survey population, which is representatively sampled from Brazil's population. Our analysis, therefore, reports how a representative sample experiences uncivil content online. Furthermore, this study focuses on one platform, while right-wing users may be more active on other platforms such as Gab and Telegram. Despite these limitations, this study provides valuable insights into the dynamics of incivility in Brazilian politics and suggests the potential issues of using automatic classifiers detecting incivility. 

Theoretically, the multidimensional conceptual framework of incivility is applicable across time periods, political cultural contexts, and platforms. More general dimensions such as IMP and HSST can also be applied beyond the political domain. However, we strongly recommend testing and retraining the classifiers when applying them to other settings, as the training samples were primarily drawn from the 2022 Brazilian Presidential Election. Additionally, other methodologies employed in this study—such as non-parametric modeling, identity annotation, and network analysis—can be also applied to other datasets containing time stamps, user profiles, and interaction data.    

To protect individual confidentiality, all datasets are securely stored and cannot be shared. The primary Python and R scripts are publicly available via the Harvard Dataverse: \url{https://doi.org/10.7910/DVN/M552GM}. Classifiers used to predict multidimensional incivility are available in the GitHub repository: \url{https://github.com/yuanzhang1227/Multidimensional_Political_Incivility_Detection}.

\section{Acknowledgments}

The authors would like to thank the editor and reviewers for their feedback. This work is supported in part by funds from the the Swiss National Science Foundation grant 100017\_204483.

\bibliography{aaai25}

\begin{thebibliography}{58}
\providecommand{\natexlab}[1]{#1}

\bibitem[{Akuma, Lubem, and Adom(2022)}]{akuma2022comparing}
Akuma, S.; Lubem, T.; and Adom, I.~T. 2022.
\newblock Comparing Bag of Words and TF-IDF with different models for hate speech detection from live tweets.
\newblock \emph{International Journal of Information Technology}, 14(7): 3629--3635.

\bibitem[{Aleksandric et~al.(2024)Aleksandric, Roy, Pankaj, Wilson, and Nilizadeh}]{aleksandric2024users}
Aleksandric, A.; Roy, S.~S.; Pankaj, H.; Wilson, G.~M.; and Nilizadeh, S. 2024.
\newblock Users’ Behavioral and Emotional Response to Toxicity in Twitter Conversations.
\newblock In \emph{Proceedings of the International AAAI Conference on Web and Social Media}, volume~18, 29--42.

\bibitem[{Badjatiya et~al.(2017)Badjatiya, Gupta, Gupta, and Varma}]{badjatiya2017deep}
Badjatiya, P.; Gupta, S.; Gupta, M.; and Varma, V. 2017.
\newblock Deep learning for hate speech detection in tweets.
\newblock In \emph{Proceedings of the 26th International Conference on World Wide Web Companion}, 759--760.

\bibitem[{Bakshy et~al.(2011)Bakshy, Hofman, Mason, and Watts}]{bakshy2011everyone}
Bakshy, E.; Hofman, J.~M.; Mason, W.~A.; and Watts, D.~J. 2011.
\newblock Everyone's an influencer: quantifying influence on Twitter.
\newblock In \emph{Proceedings of the Fourth ACM International Conference on Web Search and Data Mining}, 65--74. ACM.

\bibitem[{Bentivegna and Rega(2022)}]{bentivegna2022searching}
Bentivegna, S.; and Rega, R. 2022.
\newblock Searching for the dimensions of today’s political incivility.
\newblock \emph{Social Media+ Society}, 8(3): 20563051221114430.

\bibitem[{Brooks and Geer(2007)}]{brooks2007beyond}
Brooks, D.~J.; and Geer, J.~G. 2007.
\newblock Beyond negativity: The effects of incivility on the electorate.
\newblock \emph{American Journal of Political Science}, 51(1): 1--16.

\bibitem[{Buchinsky(1998)}]{buchinsky1998quantile}
Buchinsky, M. 1998.
\newblock Recent advances in quantile regression models: A practical guideline for empirical research.
\newblock \emph{Journal of Human Resources}, 33(1): 88--126.

\bibitem[{Chong and Kwak(2022)}]{chong2022understanding}
Chong, Y.~Y.; and Kwak, H. 2022.
\newblock Understanding toxicity triggers on Reddit in the context of Singapore.
\newblock In \emph{Proceedings of the International AAAI Conference on Web and Social Media}, volume~16, 1383--1387.

\bibitem[{Coe, Kenski, and Rains(2014)}]{coe2014online}
Coe, K.; Kenski, K.; and Rains, S.~A. 2014.
\newblock Online and uncivil? Patterns and determinants of incivility in newspaper website comments.
\newblock \emph{Journal of Communication}, 64(4): 658--679.

\bibitem[{Davidson, Sun, and Wojcieszak(2020)}]{davidson2020classifier}
Davidson, S.; Sun, Q.; and Wojcieszak, M. 2020.
\newblock Developing a New Classifier for Automated Identification of Incivility in Social Media.
\newblock In \emph{Proceedings of the Fourth Workshop on Online Abuse and Harms}, 95--101.

\bibitem[{Davidson et~al.(2017)Davidson, Warmsley, Macy, and Weber}]{davidson2017hate}
Davidson, T.; Warmsley, D.; Macy, M.; and Weber, I. 2017.
\newblock Automated Hate Speech Detection and the Problem of Offensive Language.
\newblock In \emph{Proceedings of the 11th International AAAI Conference on Web and Social Media}, 512--515.

\bibitem[{Founta et~al.(2018)Founta, Djouvas, Chatzakou, Leontiadis, Blackburn, Stringhini, and Kourtellis}]{founta2018large}
Founta, A.; Djouvas, C.; Chatzakou, D.; Leontiadis, I.; Blackburn, J.; Stringhini, G.; and Kourtellis, N. 2018.
\newblock Large scale crowdsourcing and characterization of Twitter abusive behavior.
\newblock In \emph{Proceedings of the International AAAI Conference on Web and Social Media}, volume~12.

\bibitem[{Frimer et~al.(2023)Frimer, Aujla, Feinberg, Skitka, Aquino, Eichstaedt, and Willer}]{frimer2023incivility}
Frimer, J.~A.; Aujla, H.; Feinberg, M.; Skitka, L.~J.; Aquino, K.; Eichstaedt, J.~C.; and Willer, R. 2023.
\newblock Incivility is Rising Among American Politicians on Twitter.
\newblock \emph{Social Psychological and Personality Science}, 14(2): 259--269.

\bibitem[{Gallacher, Heerdink, and Hewstone(2021)}]{gallacher2021online}
Gallacher, J.~D.; Heerdink, M.~W.; and Hewstone, M. 2021.
\newblock Online engagement between opposing political protest groups via social media is linked to physical violence of offline encounters.
\newblock \emph{Social Media+ Society}, 7(1): 2056305120984445.

\bibitem[{Gao et~al.(2024)Gao, Qin, Murali, Eckart, Zhou, Beel, and Yang}]{gao2024crisis}
Gao, Y.; Qin, W.; Murali, A.; Eckart, C.; Zhou, X.; Beel, J.~D.; and Yang, D. 2024.
\newblock A Crisis of Civility? Modeling Incivility and Its Effects in Political Discourse Online.
\newblock In \emph{Proceedings of the International AAAI Conference on Web and Social Media}, volume~18, 408--421.

\bibitem[{Gervais(2015)}]{gervais2015incivility}
Gervais, B.~T. 2015.
\newblock Incivility online: Affective and behavioral reactions to uncivil political posts in a web-based experiment.
\newblock \emph{Journal of Information Technology \& Politics}, 12(2): 167--185.

\bibitem[{Gitari et~al.(2015)Gitari, Zuping, Damien, and Long}]{gitari2015lexicon}
Gitari, N.~D.; Zuping, Z.; Damien, H.; and Long, J. 2015.
\newblock A lexicon-based approach for hate speech detection.
\newblock \emph{International Journal of Multimedia and Ubiquitous Engineering}, 10(4): 215--230.

\bibitem[{Gomez et~al.(2020)Gomez, Gibert, Gomez, and Karatzas}]{gomez2020exploring}
Gomez, R.; Gibert, J.; Gomez, L.; and Karatzas, D. 2020.
\newblock Exploring hate speech detection in multimodal publications.
\newblock In \emph{Proceedings of the IEEE/CVF Winter Conference on Applications of Computer Vision}, 1470--1478.

\bibitem[{Goodwin et~al.(2023)Goodwin, Joseff, Riedl, Lukito, and Woolley}]{goodwin2023political}
Goodwin, A.; Joseff, K.; Riedl, M.~J.; Lukito, J.; and Woolley, S. 2023.
\newblock Political relational influencers: The mobilization of social media influencers in the political arena.
\newblock \emph{International Journal of Communication}, 17: 21.

\bibitem[{Harff and Schmuck(2023)}]{harff2023influencers}
Harff, D.; and Schmuck, D. 2023.
\newblock Influencers as empowering agents? Following political influencers, internal political efficacy and participation among youth.
\newblock \emph{Political Communication}, 40(2): 147--172.

\bibitem[{Harris, Foxman, and Partin(2023)}]{harris2023dont}
Harris, B.~C.; Foxman, M.; and Partin, W.~C. 2023.
\newblock "Don't make me ratio you again": How political influencers encourage platformed political participation.
\newblock \emph{Social Media+ Society}, 9(2): 20563051231177944.

\bibitem[{Hemsley(2019)}]{hemsley2019followers}
Hemsley, J. 2019.
\newblock Followers Retweet! The Influence of Middle-Level Gatekeepers on the Spread of Political Information on Twitter.
\newblock \emph{Policy \& Internet}, 11(3): 280--304.

\bibitem[{Heseltine and Dorsey(2022)}]{heseltine2022online}
Heseltine, M.; and Dorsey, S. 2022.
\newblock Online incivility in the 2020 congressional elections.
\newblock \emph{Political Research Quarterly}, 75(2): 512--526.

\bibitem[{Hilbert et~al.(2017)Hilbert, V{\'a}squez, Halpern, Valenzuela, and Arriagada}]{hilbert2017onestep}
Hilbert, M.; V{\'a}squez, J.; Halpern, D.; Valenzuela, S.; and Arriagada, A. 2017.
\newblock One Step, Two Step, Network Step? Complementary Perspectives on Communication Flows in Twittered Citizen Protests.
\newblock \emph{Social Science Computer Review}, 35(4): 444--461.

\bibitem[{Hosseini et~al.(2017)Hosseini, Kannan, Zhang, and Poovendran}]{hosseini2017deceiving}
Hosseini, H.; Kannan, S.; Zhang, B.; and Poovendran, R. 2017.
\newblock Deceiving Google's Perspective API built for detecting toxic comments.
\newblock \emph{arXiv preprint arXiv:1702.08138}.

\bibitem[{Hunt and Gruszczynski(2024)}]{hunt2024horizontal}
Hunt, K.; and Gruszczynski, M. 2024.
\newblock “Horizontal” Two-Step Flow: The Role of Opinion Leaders in Directing Attention to Social Movements in Decentralized Information Environments.
\newblock \emph{Mass Communication and Society}, 27(2): 230--253.

\bibitem[{Katz(1957)}]{katz1957two}
Katz, E. 1957.
\newblock The two-step flow of communication: An up-to-date report on a hypothesis.
\newblock \emph{Public Opinion Quarterly}, 21(1): 61--78.

\bibitem[{Katz and Lazarsfeld(1955)}]{katz1955interpersonal}
Katz, E.; and Lazarsfeld, P. 1955.
\newblock Interpersonal Networks: Communicating within the group.
\newblock In \emph{Personal Influence}, 0--0. New York: Free Press.

\bibitem[{Khamis, Ang, and Welling(2017)}]{khamis2017self}
Khamis, S.; Ang, L.; and Welling, R. 2017.
\newblock Self-branding, ‘micro-celebrity’ and the rise of social media influencers.
\newblock \emph{Celebrity Studies}, 8(2): 191--208.

\bibitem[{Kim and Kim(2019)}]{kim2019incivility}
Kim, Y.; and Kim, Y. 2019.
\newblock Incivility on Facebook and political polarization: The mediating role of seeking further comments and negative emotion.
\newblock \emph{Computers in Human Behavior}, 99: 219--227.

\bibitem[{Lampter(2023)}]{lampter2023pink}
Lampter, M. 2023.
\newblock The Two Pink Tides in Latin America. Contemporary Global Prospects.
\newblock \emph{Human Affairs}, 33(3): 319--334.

\bibitem[{Lee(2015)}]{lee2015double}
Lee, J. 2015.
\newblock The double-edged sword: The effects of journalists' social media activities on audience perceptions of journalists and their news products.
\newblock \emph{Journal of Computer-Mediated Communication}, 20(3): 312--329.

\bibitem[{Lozano-Reich and Cloud(2009)}]{lozano2009uncivil}
Lozano-Reich, N.~M.; and Cloud, D.~L. 2009.
\newblock The uncivil tongue: Invitational rhetoric and the problem of inequality.
\newblock \emph{Western Journal of Communication}, 73(2): 220--226.

\bibitem[{Maity et~al.(2018)Maity, Chakraborty, Goyal, and Mukherjee}]{maity2018opinion}
Maity, S.~K.; Chakraborty, A.; Goyal, P.; and Mukherjee, A. 2018.
\newblock Opinion Conflicts: An Effective Route to Detect Incivility in Twitter.
\newblock \emph{Proceedings of the ACM on Human-Computer Interaction}, 2(CSCW): 1--27.

\bibitem[{Marozzo and Bessi(2018)}]{marozzo2018analyzing}
Marozzo, F.; and Bessi, A. 2018.
\newblock Analyzing polarization of social media users and news sites during political campaigns.
\newblock \emph{Social Network Analysis and Mining}, 8: 1--13.

\bibitem[{Masullo~Chen et~al.(2019)Masullo~Chen, Muddiman, Wilner, Pariser, and Stroud}]{chen2019incivility}
Masullo~Chen, G.; Muddiman, A.; Wilner, T.; Pariser, E.; and Stroud, N.~J. 2019.
\newblock We should not get rid of incivility online.
\newblock \emph{Social Media+ Society}, 5(3): 2056305119862641.

\bibitem[{Mathew et~al.(2021)Mathew, Saha, Yimam, Biemann, Goyal, and Mukherjee}]{mathew2021hatexplain}
Mathew, B.; Saha, P.; Yimam, S.~M.; Biemann, C.; Goyal, P.; and Mukherjee, A. 2021.
\newblock HateXplain: A Benchmark Dataset for Explainable Hate Speech Detection.
\newblock In \emph{Proceedings of the AAAI Conference on Artificial Intelligence}, volume~35, 14867--14875.

\bibitem[{McGregor(2020)}]{mcgregor2020taking}
McGregor, S.~C. 2020.
\newblock “Taking the temperature of the room”: how political campaigns use social media to understand and represent public opinion.
\newblock \emph{Public Opinion Quarterly}, 84(S1): 236--256.

\bibitem[{Miller(1984)}]{miller1984use}
Miller, D. 1984.
\newblock The use and abuse of political violence.
\newblock \emph{Political Studies}, 32(3): 401--419.

\bibitem[{Mollas et~al.(2022)Mollas, Chrysopoulou, Karlos, and Tsoumakas}]{mollas2022ethos}
Mollas, I.; Chrysopoulou, Z.; Karlos, S.; and Tsoumakas, G. 2022.
\newblock ETHOS: A Multi-Label Hate Speech Detection Dataset.
\newblock \emph{Complex \& Intelligent Systems}, 8(6): 4663--4678.

\bibitem[{Muddiman(2017)}]{muddiman2017personal}
Muddiman, A. 2017.
\newblock Personal and public levels of political incivility.
\newblock \emph{International Journal of Communication}, 11: 21.

\bibitem[{Mutz(2007)}]{mutz2007effects}
Mutz, D.~C. 2007.
\newblock Effects of “in-your-face” television discourse on perceptions of a legitimate opposition.
\newblock \emph{American Political Science Review}, 101(4): 621--635.

\bibitem[{Mutz and Reeves(2005)}]{mutz2005new}
Mutz, D.~C.; and Reeves, B. 2005.
\newblock The new videomalaise: Effects of televised incivility on political trust.
\newblock \emph{American Political Science Review}, 99(1): 1--15.

\bibitem[{Otto, Lecheler, and Schuck(2020)}]{otto2020context}
Otto, L.~P.; Lecheler, S.; and Schuck, A.~R. 2020.
\newblock Is context the key? The (non-) differential effects of mediated incivility in three European countries.
\newblock \emph{Political Communication}, 37(1): 88--107.

\bibitem[{Papacharissi(2004)}]{papacharissi2004democracy}
Papacharissi, Z. 2004.
\newblock Democracy Online: Civility, Politeness, and the Democratic Potential of Online Political Discussion Groups.
\newblock \emph{New Media \& Society}, 6(2): 259--283.

\bibitem[{Rega, Marchetti, and Stanziano(2023)}]{rega2023incivility}
Rega, R.; Marchetti, R.; and Stanziano, A. 2023.
\newblock Incivility in online discussion: An examination of impolite and intolerant comments.
\newblock \emph{Social Media+ Society}, 9(2): 20563051231180638.

\bibitem[{Reimers and Gurevych(2019)}]{reimers2019sentence}
Reimers, N.; and Gurevych, I. 2019.
\newblock Sentence-BERT: Sentence Embeddings using Siamese BERT-Networks.
\newblock \emph{arXiv preprint arXiv:1908.10084}.

\bibitem[{Riedl, Lukito, and Woolley(2023)}]{riedl2023political}
Riedl, M.~J.; Lukito, J.; and Woolley, S.~C. 2023.
\newblock Political Influencers on Social Media: An Introduction.
\newblock \emph{Social Media + Society}, 9(2): 20563051231177938.

\bibitem[{Rossini(2022)}]{rossini2022beyond}
Rossini, P. 2022.
\newblock Beyond Incivility: Understanding Patterns of Uncivil and Intolerant Discourse in Online Political Talk.
\newblock \emph{Communication Research}, 49(3): 399--425.

\bibitem[{Shah et~al.(2021)Shah, Udmale, Sambhe, and Bhole}]{shah2021deep}
Shah, V.; Udmale, S.~S.; Sambhe, V.; and Bhole, A. 2021.
\newblock A deep hybrid approach for hate speech analysis.
\newblock In \emph{Computer Analysis of Images and Patterns: 19th International Conference, CAIP 2021, Virtual Event, September 28–30, 2021, Proceedings, Part I}, 424--433. Springer International Publishing.

\bibitem[{Singh et~al.(2024)Singh, Ghafouri, Such, and Suarez-Tangil}]{singh2024differences}
Singh, A.~K.; Ghafouri, V.; Such, J.; and Suarez-Tangil, G. 2024.
\newblock Differences in the Toxic Language of Cross-Platform Communities.
\newblock In \emph{Proceedings of the International AAAI Conference on Web and Social Media}, volume~18, 1463--1476.

\bibitem[{Stryker, Conway, and Danielson(2016)}]{stryker2016political}
Stryker, R.; Conway, B.~A.; and Danielson, J.~T. 2016.
\newblock What is political incivility?
\newblock \emph{Communication Monographs}, 83(4): 535--556.

\bibitem[{Su and Marbach(2024)}]{su2024role}
Su, J.; and Marbach, P. 2024.
\newblock The Role of Social Support and Influencers in Social Media Communities.
\newblock \emph{arXiv preprint arXiv:2410.04619}.

\bibitem[{Theocharis et~al.(2020)Theocharis, Barber{\'a}, Fazekas, and Popa}]{theocharis2020dynamics}
Theocharis, Y.; Barber{\'a}, P.; Fazekas, Z.; and Popa, S.~A. 2020.
\newblock The dynamics of political incivility on Twitter.
\newblock \emph{Sage Open}, 10(2): 2158244020919447.

\bibitem[{Wahba(1990)}]{wahba1990spline}
Wahba, G. 1990.
\newblock \emph{Spline Models for Observational Data}.
\newblock Society for Industrial and Applied Mathematics.

\bibitem[{Wood(2017)}]{wood2017generalized}
Wood, S.~N. 2017.
\newblock \emph{Generalized Additive Models: An Introduction with R}.
\newblock Chapman and Hall/CRC.

\bibitem[{Wu et~al.(2011)Wu, Hofman, Mason, and Watts}]{wu2011whosays}
Wu, S.; Hofman, J.~M.; Mason, W.~A.; and Watts, D.~J. 2011.
\newblock Who Says What to Whom on Twitter.
\newblock In \emph{Proceedings of the 20th International Conference on World Wide Web}, 705--714.

\bibitem[{Zhang, Chen, and Lukito(2023)}]{zhang2023network}
Zhang, Y.; Chen, F.; and Lukito, J. 2023.
\newblock Network Amplification of Politicized Information and Misinformation about COVID-19 by Conservative Media and Partisan Influencers on Twitter.
\newblock \emph{Political Communication}, 40(1): 24--47.

\end{thebibliography}

\section{Paper Checklist}

\begin{enumerate}

\item For most authors...
\begin{enumerate}
    \item  Would answering this research question advance science without violating social contracts, such as violating privacy norms, perpetuating unfair profiling, exacerbating the socio-economic divide, or implying disrespect to societies or cultures?
    \answerYes{Yes}
  \item Do your main claims in the abstract and introduction accurately reflect the paper's contributions and scope?
    \answerYes{Yes}
   \item Do you clarify how the proposed methodological approach is appropriate for the claims made? 
    \answerYes{Yes}
   \item Do you clarify what are possible artifacts in the data used, given population-specific distributions?
    \answerYes{Yes}
  \item Did you describe the limitations of your work?
    \answerYes{Yes}
  \item Did you discuss any potential negative societal impacts of your work?
    \answerYes{Yes}
      \item Did you discuss any potential misuse of your work?
    \answerYes{Yes}
    \item Did you describe steps taken to prevent or mitigate potential negative outcomes of the research, such as data and model documentation, data anonymization, responsible release, access control, and the reproducibility of findings?
    \answerYes{Yes}
  \item Have you read the ethics review guidelines and ensured that your paper conforms to them?
    \answerYes{Yes}
\end{enumerate}

\item Additionally, if your study involves hypotheses testing...
\begin{enumerate}
  \item Did you clearly state the assumptions underlying all theoretical results?
    \answerNA{NA}
  \item Have you provided justifications for all theoretical results?
    \answerNA{NA}
  \item Did you discuss competing hypotheses or theories that might challenge or complement your theoretical results?
    \answerNA{NA}
  \item Have you considered alternative mechanisms or explanations that might account for the same outcomes observed in your study?
    \answerNA{NA}
  \item Did you address potential biases or limitations in your theoretical framework?
    \answerNA{NA}
  \item Have you related your theoretical results to the existing literature in social science?
    \answerNA{NA}
  \item Did you discuss the implications of your theoretical results for policy, practice, or further research in the social science domain?
    \answerNA{NA}
\end{enumerate}

\item Additionally, if you are including theoretical proofs...
\begin{enumerate}
  \item Did you state the full set of assumptions of all theoretical results?
    \answerNA{NA}
	\item Did you include complete proofs of all theoretical results?
    \answerNA{NA}
\end{enumerate}

\item Additionally, if you ran machine learning experiments...
\begin{enumerate}
  \item Did you include the code, data, and instructions needed to reproduce the main experimental results (either in the supplemental material or as a URL)?
    \answerNo{No,
the data is collected from human subjects and could
potentially be used to identify personal information.
Therefore, only the code and instructions are provided}
  \item Did you specify all the training details (e.g., data splits, hyperparameters, how they were chosen)?
    \answerYes{Yes}
     \item Did you report error bars (e.g., with respect to the random seed after running experiments multiple times)?
    \answerNo{No, the results of classification of our study are not
significantly influenced by randomization.}
	\item Did you include the total amount of compute and the type of resources used (e.g., type of GPUs, internal cluster, or cloud provider)?
    \answerNo{No, all the machine learning models were run on a local machine without the use of GPUs, internal clusters, or cloud resources.}
     \item Do you justify how the proposed evaluation is sufficient and appropriate to the claims made? 
    \answerYes{Yes}
     \item Do you discuss what is ``the cost`` of misclassification and fault (in)tolerance?
    \answerYes{Yes}
  
\end{enumerate}

\item Additionally, if you are using existing assets (e.g., code, data, models) or curating/releasing new assets, \textbf{without compromising anonymity}...
\begin{enumerate}
  \item If your work uses existing assets, did you cite the creators?
    \answerNA{NA}
  \item Did you mention the license of the assets?
    \answerNA{NA}
  \item Did you include any new assets in the supplemental material or as a URL?
    \answerNA{NA}
  \item Did you discuss whether and how consent was obtained from people whose data you're using/curating?
    \answerNA{NA}
  \item Did you discuss whether the data you are using/curating contains personally identifiable information or offensive content?
    \answerNA{NA}
\item If you are curating or releasing new datasets, did you discuss how you intend to make your datasets FAIR ?
\answerNA{NA}
\item If you are curating or releasing new datasets, did you create a Datasheet for the Dataset? 
\answerNA{NA}
\end{enumerate}

\item Additionally, if you used crowdsourcing or conducted research with human subjects, \textbf{without compromising anonymity}...
\begin{enumerate}
  \item Did you include the full text of instructions given to participants and screenshots?
    \answerYes{Yes}
  \item Did you describe any potential participant risks, with mentions of Institutional Review Board (IRB) approvals?
    \answerYes{Yes}
  \item Did you include the estimated hourly wage paid to participants and the total amount spent on participant compensation?
    \answerNo{No, our survey was conducted by a sur-
vey company, so we did not pay the participants di-
rectly.}
   \item Did you discuss how data is stored, shared, and deidentified?
   \answerYes{Yes}
\end{enumerate}

\end{enumerate}

\vspace{2em}

\appendix
\section{Appendix}
\subsection{Survey Sample validation}


\begin{figure}[H]
\centering
\includegraphics[width=1.0\linewidth]{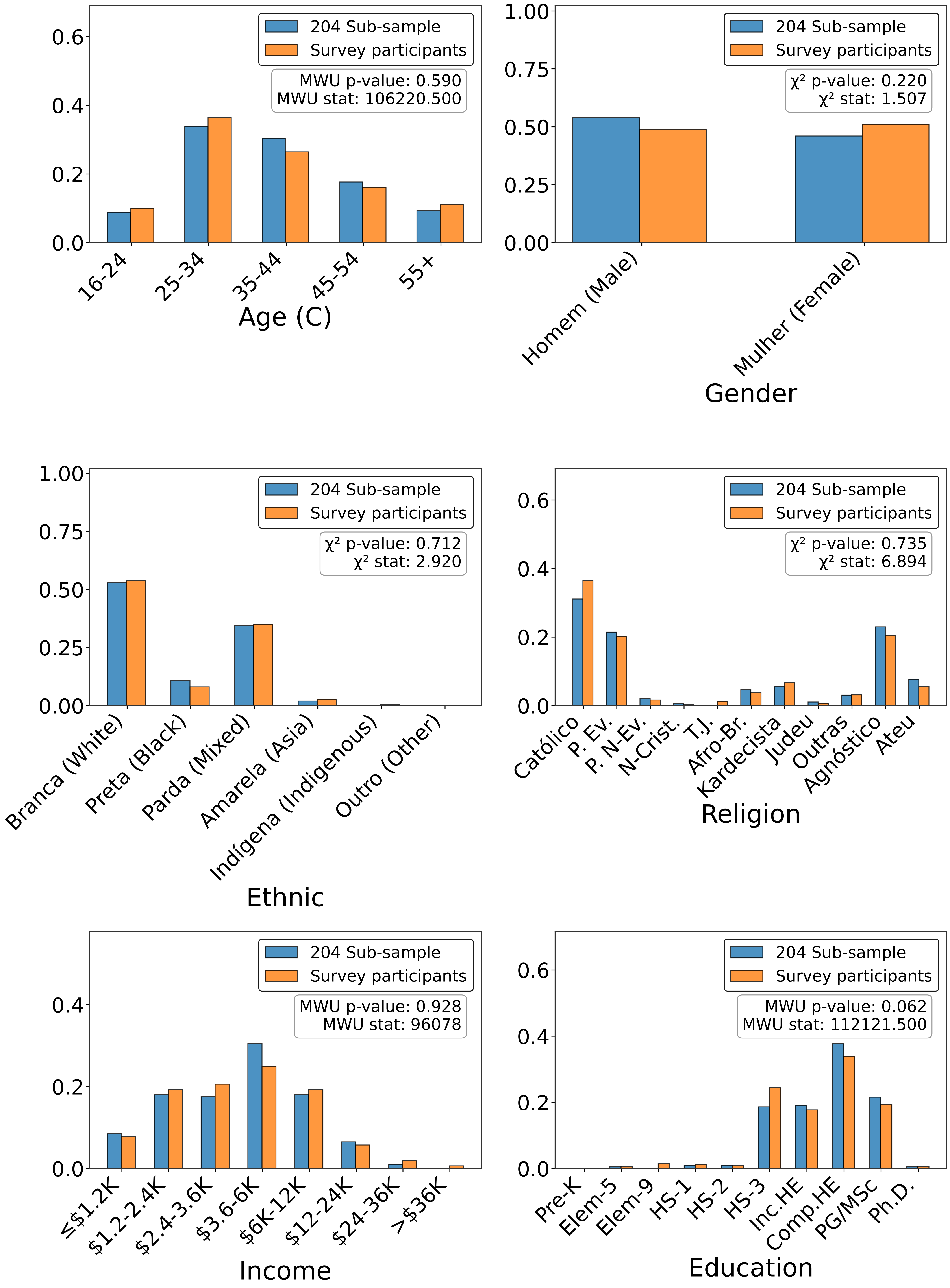}
\caption{Comparison of distributions on Age, Gender, Ethnic, Religion, Income, and Education between survey respondents (N = 1,018) and its sub-sample (N = 204).}
\label{fig:sample validation}
\end{figure}

\begin{figure}[H]
\centering
\includegraphics[width=1.0\linewidth]{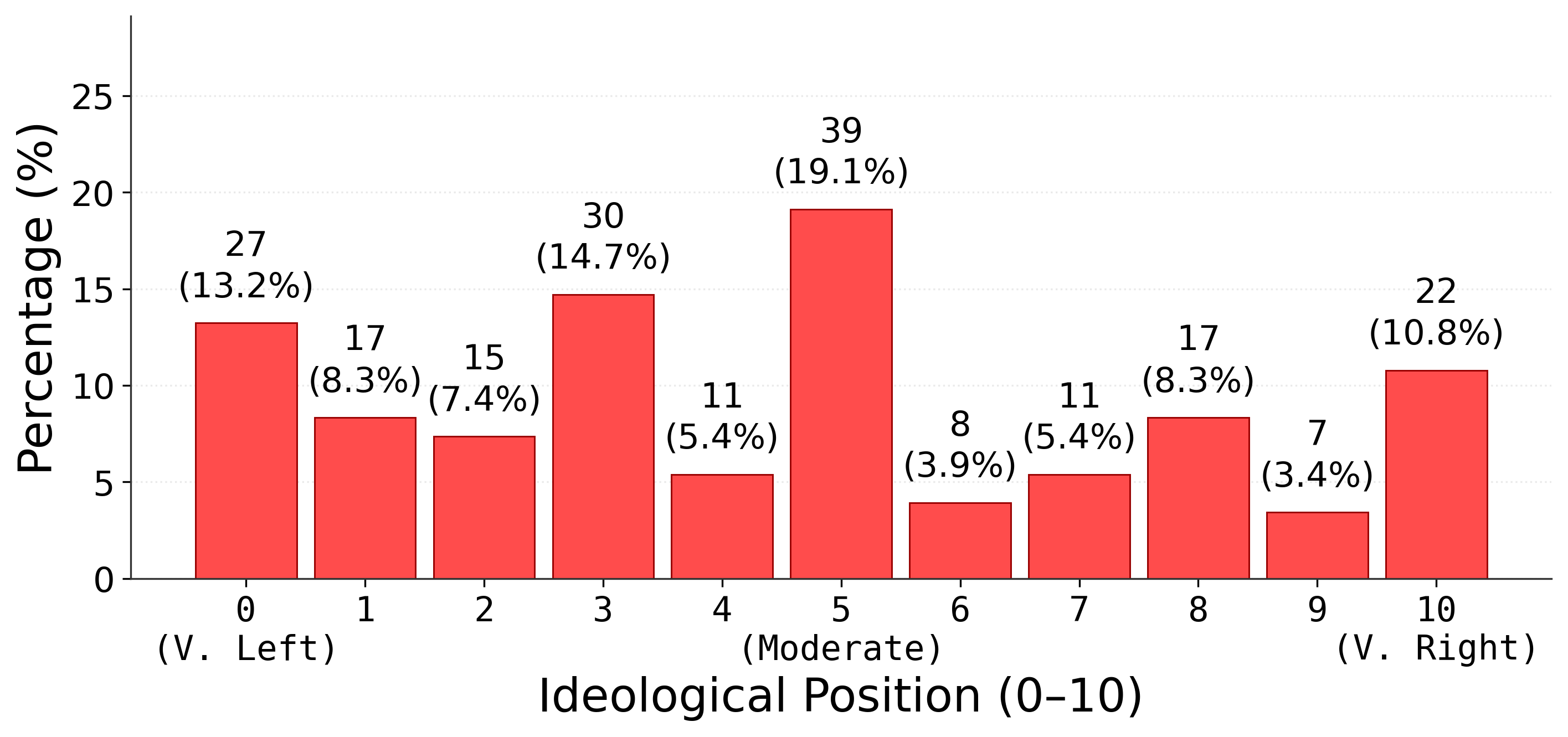}
\caption{Distribution of ideological position among the survey sub-sample (N = 204).}
\label{fig:ideology distribution}
\end{figure}

\subsection{Identifcation of Political Influencers}

\begin{figure}[!b]
\centering
\includegraphics[width=0.9\linewidth]{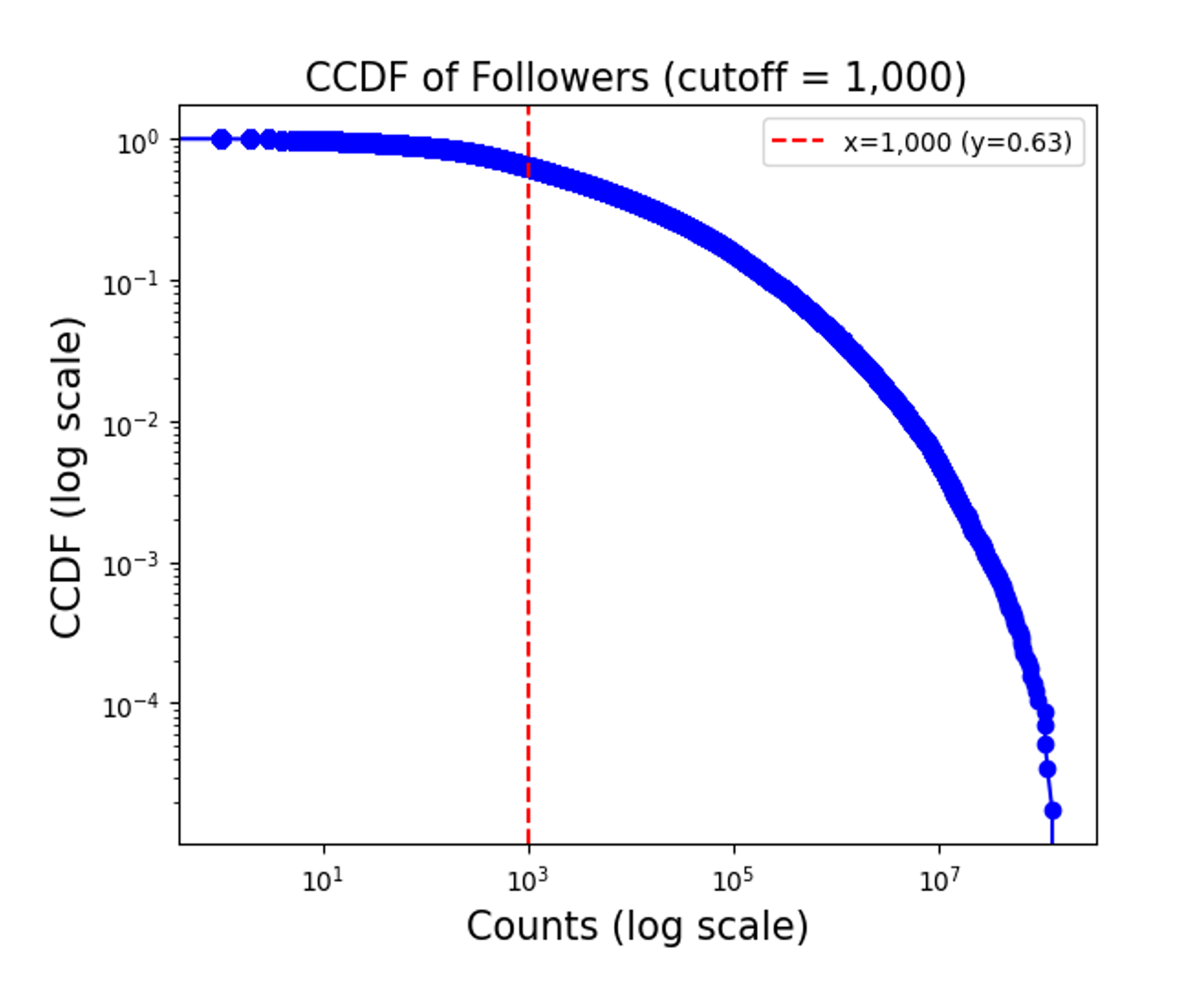}
\caption{The complementary cumulative distribution function (CCDF) plot of the number of followers of Twitter/X accounts followed by survey respondents, with cutoffs at 1,000.}
\label{fig:CCDF}
\end{figure}

We use a heuristic strategy of identifying political influencers from the 57,645 accounts followed by survey respondents. We define political influencers as a composition of both ordinary citizens and celebrities (e.g., politicians, parties, media outlets, journalists, and individuals) who satisfy two conditions: 1) are influential and 2) are likely to produce political content. According to this definition, we select political influencers in three steps. Firstly, we identify influential accounts that have a number of followers exceeding 1,000. Fig. \ref{fig:CCDF} displays the Complementary Cumulative Distribution Function (CCDF) plot for the number of followers of accounts followed by survey respondents. We establish a threshold of 1,000 followers, and only accounts exceeding this threshold are retained, accounting for 63\% of the total accounts who are followed by survey respondents.

Second, from the accounts with more than 1,000 followers, we further select Brazilian accounts based on the location information displayed in their profile (Brazil or Brazilian cities). Third, we filter the accounts that might produce political information from the Brazilian influencers, covering categories of politicians, parties, media outlets, journalists, and individuals. We manually examine approximately 2,000 random profile examples and create a politically relevant keyword list based on these samples. Accounts potentially generating political content are identified by matching politics-related keywords in their profiles and supplemented by additional lists of politicians (based on 2022 presidential election candidates), parties, and media outlets (based on Digital News Report 2022 produced by Reuters Institute Oxford). See Tab. \ref{table:political_influencers} for more details.

\begin{table*}[htbp]
\centering
\begin{tabular}{p{2cm} p{3cm} p{10cm}}
\toprule
\textbf{Criteria} & \textbf{Category} & \textbf{Keywords} \\
\midrule
\makecell[l]{\textbf{Political} \\ \textbf{Keywords}}
 & General & política, político, political, politics, democracia, democracy \\
                           & Election & bolsonaro, bolsonarista, lula, lulista, candidato, partido, presidente \\
                           & Public sector & federal, conselho nacional de, ministro, senador, deputado, governador, prefeito, vereador, secretário \\
                           & Ideology & conservador, conservative, liberal, liberalismo, libertairia, esquerdopata, esquerda, direita, direitista, comunista, comunismo, nacionalista, patriota, globalista, feminista, armamentista, fascista, racist, colonialista, socialista, ativista, progressista \\
                           & Topic (culture) & aborto, mulher, preta, lgbt, gay, bissexualismo, homophobic, catílico, jesus, deus, ambiente, clima, justiça, imigrante, foreigner \\
                           & Topic (economic) & economia, bem-estar, pobre, desigualdade \\
\midrule
\makecell[l]{\textbf{Political} \\ \textbf{Accounts}}
 & Political party & Partido da Mulher Brasileira, Partido dos Trabalhadores, Partido da Social Democracia Brasileira, Progressistas, Partido Democrático Trabalhista, Partido Trabalhista Brasileiro, União Brasil, Partido Liberal, Partido Socialista Brasileiro, Republicanos, Cidadania, Partido Comunista do Brasil, Partido Social Cristão, Podemos, Partido Social Democrático, Partido Verde, Patriota, Solidariedade, Partido da Mobilização Nacional, Avante, Partido Trabalhista Cristão, Partido Socialismo e Liberdade, Democracia Cristã, Partido Renovador, Trabalhista Brasileiro, Partido Republicano da Ordem Social, Partido da Mulher Brasileira, Partido Novo, Rede Sustentabilidade, Partido Socialista dos Trabalhadores Unificado, Partido Comunista Brasileiro, Partido da Causa Operária, Unidade Popular, Avante, Agir, MDB Nacional \\
                           & Politician & Aldo Rebelo, Soraya Thronicke, Jair Bolsonaro, Luiz Inácio Lula da Silva, Ciro Gomes, Simone Tebet, André Janones, Luiz Felipe D'Avila, José Maria Eymael, Leonardo Péricles, Sofia Manzano, Vera Lúcia Salgado, Luciano Bivar, Pablo Marçal, Wilson Witzel, Janaina Paschoal, José Reguff, Ibaneis Rocha, Renan Filho, Renato Casagrande, Michel Temer, Jorge Kajuru, Padre Kelmon \\
\midrule
\makecell[l]{\textbf{Media} \\ \textbf{Keywords}}
 & Individual aggregator & jornalista, journalist, correspondent, repórter, comandante, commentator, comentarista, influencer, news, semanal \\
\midrule
\makecell[l]{\textbf{Media} \\ \textbf{Accounts}}
 & News outlet & Globo News online (incl. G1), UOL online, Record News online (incl. R7.com), O Globo online, Band News online, Folha de S. Paulo online, O Estado de S. Paulo online, BBC News online, Rede TV News online, notícias, Jornal Extra online, TV SBT (incl. SBT Brasil), TV Band News, CNN, TV Brasil (public broadcaster) \\
\bottomrule
\end{tabular}
\caption{Identification of Political Influencers}
\label{table:political_influencers}
\end{table*}

The three steps result in identifying 2,307 Brazilian political influencers from the 57,645 followed accounts.

\subsection*{Codebook for Socio-Political Identity Annotation}

This codebook is designed to guide the systematic annotation of socio-political attributes of Brazilian Twitter/X political influencers, with a focus on account type, ideological position, campaign support, and social identity. It provides detailed instructions for coding five key dimensions: (1) whether the user potentially produces political content, (2) the type of account (e.g., politician, media, individual), (3) the user's ideological position, if declared, (4) explicit support for political candidates in the 2022 Brazilian Presidential Election (Lula or Bolsonaro), and (5) any self-disclosed or publicly visible social identities (e.g., Women, Black, LGBTQ, Religious). These annotations support the analysis of uncivil political discourse on Brazilian Twitter/X.

\vspace{1em}

\textbf{COL 1: Politics} \\
\textbf{Options}: Yes / No

\textbf{Criteria:} \\
Code \textbf{Yes} if any of the following conditions are met:
\begin{enumerate}
    \item The text is posted by a Brazilian politician, political party, other political agencies, media outlets, or media workers such as journalists/reporters/commenters/columnists (note: media outlets or media workers for sports/music/fashion or other non-political industries do not count).
    \item The text include any words related to ideological position leaning (e.g., left, right, liberal, conservative etc.), campaign or social movements slogans (e.g., "VoltaLula," "ForaBolsonaro," "Eleicoes2022"), or mentions political/social issues (e.g., "social welfare", "environmental policy", "abortion rights", "minority rights").
\end{enumerate}
Otherwise, code \textbf{No}.

\vspace{1em}
If the Politics column is marked 'Yes', assign the following labels accordingly:
\vspace{1em}

\textbf{COL 2: Account Type} \\
\textbf{Options}: Politician / Media / Individual

\textbf{Criteria:}
\begin{itemize}
    \item Code \textbf{Politician} if the user is a Brazilian politician.
    \item Code \textbf{Media} if the user is a media outlet or a media worker such as journalist/reporter/commenter/columnist.
    \item Code \textbf{Individual} if the user is an ordinary user, including celebrities, scholars, and activists.
\end{itemize}

\vspace{1em}

\textbf{COL 3: Ideological Position} \\
\textbf{Options}: Left / Right / Center

\textbf{Criteria:}
\begin{itemize}
    \item \textbf{For Politicians}, assign ideology based on party affiliation:
    \begin{itemize}
        \item \textbf{Left} for PT, PSOL, PCdoB, PDT, PSB, and other left-wing parties
        \item \textbf{Right} for PL, NOVO, and other right-wing parties
        \item \textbf{Center} for centrist parties (e.g., MDB, PSD)
    \end{itemize}
    \item \textbf{For Media}, refer to known political orientation based on public evaluations (e.g., Media Bias/Fact Check). Leave blank if unclear.
    \item \textbf{For Individuals}, code based on explicit ideological declaration in the profile:
    \begin{itemize}
        \item \textbf{Left}: mentions being left, liberal, or supports left-wing agendas
        \item \textbf{Right}: mentions being right, conservative, or supports right-wing agendas
        \item \textbf{Center}: declares that they are at central position (e.g., neither left nor right)
        \item Leave blank if no ideological position is stated explicitly.
    \end{itemize}
\end{itemize}

\vspace{1em}

\textbf{COL 4: Campaign Support} \\
\textbf{Options}: Lula camp / Bolsonaro camp

\textbf{Criteria:}
\begin{itemize}
    \item Code \textbf{Lula camp} if the user is a politician and endorsed Lula during the 2022 Brazilian Presidential Election, or an Individual indicating support Lula or against Bolsonaro (e.g., ForaBolsonaro).
    \item Code \textbf{Bolsonaro camp} if the user is a politician and endorsed Bolsonaro during the 2022 Brazilian Presidential Election, or an Individual indicating support Bolsonaro or against Lula (e.g., LulaLadrão).
    \item If neither support nor opposition is clearly indicated, leave blank.
\end{itemize}

\vspace{1em}

\textbf{COL 5: Social Identity} \\
\textbf{Options}: Women / Black / LGBTQ / Religious

\textbf{Criteria:}
\begin{itemize}
    \item Code \textbf{Women} if the user represents image for the women group (e.g., advocate for women; other female titles
such as mom, girl, grandma do not count) or supports feminism.
    \item Code \textbf{Black} if the user self-identifies as Black or advocates for Black rights.
    \item Code \textbf{LGBTQ} if the user self-identifies as LGBTQ (e.g., gay, lesbian, trans, bi) or supports LGBTQ rights.
    \item Code \textbf{Religious} if the user expresses religious affiliation (e.g., Christian, Evangelical, Catholic) or references faith-based communities, ministries, or biblical quotes.
    \item Use multiple labels if applicable, e.g., (Women, Black).
    \item If no identity is clearly indicated, leave blank.
\end{itemize}

Please note that not all political influencers disclose their identities, we assign those unrevealed profiles as ``Unlabeled''.

\subsection{Codebook for Multidimentional Incivility Annotation}

The training samples for the automatic classifier are drawn from a corpus compiled by the Swiss National Science Foundation (SNSF) project \textit{From Uncivil Disagreement to Political Unrest? A Cross-Platform \& Cross-National Analysis of the Offline Consequences of Online Incivility}. 
The dataset includes posts from media outlets, political candidates, and political parties across multiple platforms—Twitter/X, Facebook, YouTube, Telegram, etc.—with a detailed source list provided in Tab. \ref{tab:source_list}. Media outlets are selected based on the most popular sources provided in the Digital News Report 2022, while political candidates and parties include those who participated in the 2022 Brazilian Presidential Election. Comments on these posts are also collected.

Training samples, including both uncivil (positive) and civil (negative) examples, are selected using the Perspective API in combination with stratified random sampling—stratified by platform and user account—for manual annotation by human coders.

The following codebook guides the annotation process across four dimensions: Impoliteness, Physical Harm and Violent Political Rhetoric, Hate Speech and Stereotyping, and Threats to Democratic Institutions and Values. This codebook is also part of the project. It defines political incivility across multiple dimensions and provides examples to guide coders in annotation. \\

\begin{table*}[htbp]
\centering
\begin{tabular}{p{5cm} p{3cm} p{5cm} p{3cm}}
\toprule
\textbf{Account} & \textbf{Type} & \textbf{Account} & \textbf{Type} \\
\midrule

\multicolumn{4}{l}{\textit{Media Outlets}} \\
Globo News online (incl. G1) & Media Outlet & UOL online & Media Outlet \\
Record News online (incl. R7.com) & Media Outlet & O Globo online & Media Outlet \\
Band News online & Media Outlet & Folha de S. Paulo online & Media Outlet \\
O Estado de S. Paulo online & Media Outlet & BBC News online & Media Outlet \\
Rede TV News online & Media Outlet & Jornal Extra online & Media Outlet \\
TV SBT (incl. SBT Brasil) & Media Outlet & TV Band News & Media Outlet \\
CNN & Media Outlet & TV Brasil (public broadcaster) & Media Outlet \\

\addlinespace[0.5em]
\multicolumn{4}{l}{\textit{Political Candidates}} \\
Aldo Rebelo & Political Candidate & Jair Bolsonaro & Political Candidate \\
Soraya Thronicke & Political Candidate & Luiz Inácio Lula da Silva & Political Candidate \\
Ciro Gomes & Political Candidate & Simone Tebet & Political Candidate \\
André Janones & Political Candidate & Luiz Felipe D'Avila & Political Candidate \\
José Maria Eymael & Political Candidate & Leonardo Péricles & Political Candidate \\
Sofia Manzano & Political Candidate & Vera Lúcia Salgado & Political Candidate \\
Luciano Bivar & Political Candidate & Pablo Marçal & Political Candidate \\
Wilson Witzel & Potential Candidate & Janaina Paschoal & Potential Candidate \\
José Reguff & Potential Candidate & Ibaneis Rocha & Potential Candidate \\
Renan Filho & Potential Candidate & Renato Casagrande & Potential Candidate \\
Michel Temer & Potential Candidate & Jorge Kajuru & Potential Candidate \\
Padre Kelmon & Potential Candidate & & \\

\addlinespace[0.5em]
\multicolumn{4}{l}{\textit{Political Parties}} \\
Partido da Mulher Brasileira & Political Party & Partido dos Trabalhadores & Political Party \\
Partido da Social Democracia Brasileira & Political Party & Progressistas & Political Party \\
Partido Democrático Trabalhista & Political Party & Partido Trabalhista Brasileiro & Political Party \\
União Brasil & Political Party & Partido Liberal & Political Party \\
Partido Socialista Brasileiro & Political Party & Republicanos & Political Party \\
Cidadania & Political Party & Partido Comunista do Brasil & Political Party \\
Partido Social Cristão & Political Party & Podemos & Political Party \\
Partido Social Democrático & Political Party & Partido Verde & Political Party \\
Patriota & Political Party & Solidariedade & Political Party \\
Partido da Mobilização Nacional & Political Party & Avante & Political Party \\
Partido Trabalhista Cristão & Political Party & Partido Socialismo e Liberdade & Political Party \\
Democracia Cristã & Political Party & Partido Renovador Trabalhista Brasileiro & Political Party \\
Partido Republicano da Ordem Social & Political Party & Partido Novo & Political Party \\
Rede Sustentabilidade & Political Party & Partido Socialista dos Trabalhadores Unificado & Political Party \\
Partido Comunista Brasileiro & Political Party & Partido da Causa Operária & Political Party \\
Unidade Popular & Political Party & Agir & Political Party \\
MDB Nacional & Political Party & & \\
\bottomrule
\end{tabular}
\caption{List of Sources for Training Data}
\label{tab:source_list}
\end{table*}

\vspace{1em}

\textbf{DIM 1: Impoliteness} \\

\textbf{Definition:} Here we code any kind of rudeness and disrespect, which can be directed by means of offensive language against any kind of person or group. This is meant to be understood as impoliteness in general, it does not have to be politically motivated. By impoliteness and disrespect, we understand more specifically:
\begin{itemize}
    \item Name-calling (e.g., "weirdo," "traitor," "crackpot," "thieves")
    \item Aspersions (e.g., "reckless," "stupid," "irrational," "un-American")
    \item Synonyms for lying (e.g., "hoax," "farce")
    \item Hyperbole (e.g., "outrageous," "heinous")
    \item Words that indicate non-cooperation (e.g., "polarized," "filibuster," "inflexible")
    \item Pejorative speak (e.g., "bellyache," "doublespeak", "gibberish")
    \item Vulgarity (e.g., "damn," "shit," "hell," "assholes")
    \item Belittling others
    \item Using all-caps or excessive exclamation marks to imitate shouting or screaming (e.g., “a MILLION social parasites,” “Send them back!!!!”) (context dependent)
\end{itemize}

\textbf{Operationalization:}\\
1 = Present: Message contains impoliteness.\\
0 = Not Present: No impoliteness.\\

\textbf{DIM 2: Physical Harm and Violent Political Rhetoric} \\

\textbf{Definition:} 

\begin{itemize}
    \item Messages threatening physical harm against political actors or inciting others to inflict harm. Includes direct expressions of intent to cause physical harm or indirect expressions (metaphorically speaking). The rule of thumb of classifying this dimension is to check whether there is an intention of causing actual physical harm, and it should be combined with specific contexts. 
    \item Advocating for violence as a means to achieve political ends, or suggesting that violent acts against certain individuals or groups are justified. 
\end{itemize}

\textbf{Examples:}
\begin{itemize}
    \item “I will kill you” (direct violence). Or: “hang @MikePence!” 
    \item "The only way to get things done is to take it to the streets and make them listen, by force if necessary."
\end{itemize}

PLEASE NOTE: 
\begin{itemize}
    \item Do not code this category if the underlying meaning does not describe violence (for instance in ‘get his ass kicked’ or ‘burn in hell’ – these are ‘impolite’ expressions of anger) or if the phrase is meant ironically or sarcastically (‘why don’t you shoot them all if you believe violence solves any problems’), or if someone’s violent rhetoric is just quoted. Only code if there is a clear intention of violence.
\end{itemize}

\textbf{Operationalization:}\\
1 = Present: Contains physical harm or violent political rhetoric.\\
0 = Not Present: No physical harm or violent political rhetoric.\\

\textbf{DIM 3: Hate Speech and Stereotyping} \\

\textbf{Definition:} Messages including discriminatory statements against individuals and groups who are attributed negative stereotypes based on gender identity, sexual orientation, religious beliefs, race, nationality, ideology, or disability. Messages often include plural forms and imply references to groups based on social identity.
\begin{itemize}
    \item Messages that are misogynist, xenophobic, sexist, racist, …
    \item Depicting people negatively as members of an outgroup or pariah group. 
    \item Targeted criticism based on an individual’s personality, appearance, or looks.
    \item Making over-generalizing assumptions about thoughts or behaviors of groups or individuals based on stereotypes.
\end{itemize}

\textbf{Examples:}\\
\begin{itemize}
    \item “Muslims are terrorist sympathizers.” 
    \item “Gun-owners/supporters are paranoid.”
    \item "Liberals are less patriotic."
    \item "Immigrants rely on social benefits."
    \item "Women are poor drivers."
\end{itemize}

PLEASE NOTE: 
\begin{itemize}
    \item Social identity-based groups include racial and ethnic communities, religious groups, LGBTQ+ communities, women, people with disabilities, or any other group defined by shared social attributes. Promoting tolerance, respect, and understanding is crucial in fostering a more inclusive and equitable society.
    \item Rare cases where politicians are attacked as members of these groups (as women, as blacks, etc) should also be coded here.
    \item Political ideology can also become relevant here. Clearly discriminatory, overgeneralizing pejoratives against people as Nazis, fascists, gun-toting conservatives, or as communists, leftist scum, woke snowflakes would also be coded here.
    \item Positive or neutral stereotyping is never coded.
\end{itemize}

\textbf{Operationalization:}\\
1 = Present: Message contains at least one instance of hate speech and stereotyping.\\
0 = Not Present: Message does not exhibit hate speech and stereotyping.\\

\textbf{DIM 4: Threats to Democratic Institutions and Values} \\

\textbf{Definition:} Messages undermining democratic procedures and institutions, the democratic state, and democratic values. 

\begin{itemize}
    \item Messages promoting force against the government or the forceful replacement of the existing government, resorting to violence in coup or revolution.
    
    \item Promoting Autocracy. Messages that argue in favor of undemocratic forms of governance, such as autocracy or dictatorship, over democratic principles.
    \item Discrediting democratic institutions. Messages that aim to delegitimize or undermine the importance, role, or integrity of key democratic institutions. For instance, suggesting that elections are rigged without any substantive proof, or consistently attacking the judiciary or the media.
    \item Discrediting democratic values. This includes discrediting Freedom of Speech (to express ideas without censorship or fear), Equality (that everyone has same legal rights and opportunities), Rule of Law (that everyone must abide by the law, including those in power), Free and Fair Elections (to choose your representatives without undue influence or discrimination), Civil Liberties (protecting individuals from arbitrary government interference), Pluralism (acceptance of diverse viewpoints), Accountability (elected officials and institutions are answerable to the public for their actions).
\end{itemize}

\textbf{Examples:}\\
\begin{itemize}
    \item "We don't need elections, we need a strong leader who knows what's best."

    \item "The entire electoral process is a sham. Our votes don't matter." 
\end{itemize}

\textbf{Operationalization:}\\
1 = Present: Message contains at least one instance of threats to democratic institutions and values.\\
0 = Not Present: Message does not exhibit threats to democratic institutions and values.

\end{document}